\documentclass[preprint,11pt,authoryear,a4paper]{elsarticle}

\usepackage[utf8]{inputenc}
\usepackage{amssymb}
\usepackage{amsmath}
\usepackage{amsfonts}
\usepackage{array,arydshln}
\usepackage{booktabs} 
\usepackage{multirow}
\usepackage[figuresright]{rotating}
\usepackage{pdflscape}
\usepackage{natbib}
\usepackage{color}
\usepackage{pifont}
\usepackage{float}
\usepackage[hypcap]{caption}
\usepackage{textcomp}
\usepackage{calc}
\usepackage{setspace}
\usepackage{enumitem}
\usepackage{xfrac}
\usepackage{makecell}

\usepackage[rm={lining=true}]{cfr-lm}
\usepackage{algorithm}
\usepackage{algpseudocode}
\usepackage{float}
\floatstyle{plaintop}
\newfloat{myalgorithm}{!tbp}{lop}
\floatname{myalgorithm}{Algorithm}
\usepackage{bm}	

\newcommand{\HRule}{\rule{\linewidth}{0.08em}}

\definecolor{colorlink}{rgb}{0, 0, .6}  
\definecolor{darkgreen}{rgb}{0, .5, 0}  
\definecolor{orange}{rgb}{.7, .40, 0}   
\definecolor{grey}{rgb}{.6, .6, .6}     

\newcommand{\lspac}{1.20}  

\usepackage[pdftex]{hyperref}
\hypersetup{colorlinks=true,citecolor=colorlink,filecolor=colorlink,linkcolor=colorlink,urlcolor=colorlink}

\journal{NeuroImage}

\begin{document}

\begin{frontmatter}

\title{\textbf{Permutation Inference for\\Canonical Correlation Analysis}}

\author[nih]{Anderson M.\ Winkler\corref{corresponding}}
\ead{anderson.winkler@nih.gov}
\cortext[corresponding]{Corresponding author.}
\author[ug]{Olivier Renaud}
\author[win]{\\Stephen M.\ Smith}
\author[big]{Thomas E.\ Nichols}

\address[nih]{National Institute of Mental Health (\textsc{nimh}),\\National Institutes of Health (\textsc{nih}), Bethesda, \textsc{md}, \textsc{usa}}
\address[ug]{Methodology and Data Analysis, Department of Psychology, \\University of Geneva, Switzerland}
\address[win]{Wellcome Trust Centre for Integrative Neuroimaging (\textsc{win-fmrib}),\\University of Oxford, Oxford, \textsc{uk}}
\address[big]{Big Data Institute, University of Oxford, Oxford, \textsc{uk}}


\begin{abstract}
Canonical correlation analysis (CCA) has become a key tool for population neuroimaging, allowing investigation of associations between many imaging and non-imaging measurements. As other variables are often a source of variability not of direct interest, previous work has used CCA on residuals from a model that removes these effects, then proceeded directly to permutation inference. We show that such a simple permutation test leads to inflated error rates. The reason is that residualisation introduces dependencies among the observations that violate the exchangeability assumption. Even in the absence of nuisance variables, however, a simple permutation test for CCA also leads to excess error rates for all canonical correlations other than the first. The reason is that a simple permutation scheme does not ignore the variability already explained by previous canonical variables. Here we propose solutions for both problems: in the case of nuisance variables, we show that transforming the residuals to a lower dimensional basis where exchangeability holds results in a valid permutation test; for more general cases, with or without nuisance variables, we propose estimating the canonical correlations in a stepwise manner, removing at each iteration the variance already explained, while dealing with different number of variables in both sides. We also discuss how to address the multiplicity of tests, proposing an admissible test that is not conservative, and provide a complete algorithm for permutation inference for CCA.
\end{abstract}

\begin{keyword}
canonical correlation analysis, permutation test, closed testing procedure. 
\end{keyword}
\end{frontmatter}

\renewcommand\floatpagefraction{.001}
\makeatletter
\setlength\@fpsep{\textheight}
\makeatother

\vspace{1cm}

\section{Introduction}
\label{sec:intro}
\setstretch{\lspac}

\emph{Canonical correlation analysis} (\textsc{cca}) \citep{Jordan1875, Hotelling1936} is a multivariate method that aims at reducing the correlation structure between two sets of variables to the simplest possible form (hence the name ``canonical'') through linear transformations of the variables within each set. Put simply, given two sets of variables, the method seeks linear mixtures within each set, such that each resulting mixture from one set is maximally correlated with a corresponding mixture from the other set, but uncorrelated with all other mixtures in either set.

From a peak use through from the late 1970's until mid-1980's, the method has recently regained popularity, presumably thanks to its ability to uncover latent, common factors underlying association between multiple measurements obtained, something relevant in recent research, particularly in the field of brain imaging, that uses high dimensional phenotyping and investigates between-subject variability across multiple domains. This is in contrast to initial studies that introduced \textsc{cca} to the field \citep{Friston1995, Friston1996, Worsley1997, Friman2001, Friman2002, Friman2003} for investigation of signal variation in functional magnetic resonance imaging (f\textsc{mri}) time series. For example, \citet{Smith2015} used \textsc{cca} to identify underlying factors associating brain connectivity features to various demographic, psychometric, and lifestyle measures; \citet{Rosa2015} used a sparse \textsc{cca} method to investigate differences in brain perfusion after administration of two distinct antipsychotic drugs; \citet{Miller2016} used \textsc{cca} to identify associations between imaging and non-imaging variables in the \textsc{uk} Biobank; \citet{Drysdale2017} used \textsc{cca} to investigate associations between brain connectivity and clinical assessments, and found two canonical variables that would allow classification of participants into distinct categories \citep[but see][]{Dinga2019}; \citet{Kernbach2018} used \textsc{cca} to identify connectivity patterns in the default mode network associated with patterns of connectivity elsewhere in the brain; \citet{Bijsterbosch2018}, \citet{Xia2018}, and \citet{Mihalik2019} likewise used \textsc{cca} to identify associations between functional connectivity and various indices of behaviour and psychopathology, whereas \citet{Sui2018} used a combination of multivariate methods, including \textsc{cca}, to investigate brain networks associated with composite cognitive scores; \citet{Li2019} used \textsc{cca} to investigate, among subjects, the topography of the global f\textsc{mri} signal and its relationship with a number of cognitive and behavioral measurements; \citet{Ing2019} used \textsc{cca} to identify symptom groups that were correlated with brain regions assessed through a diverse set of imaging modalities; \citet{Alnaes2020} used \textsc{cca} to investigate the association between imaging measurements and cognitive, behavioral, psychosocial and socioeconomic indices; \citet{Clemens2020} used a combination of pattern classification algorithms and \textsc{cca} to study imaging and behavioral correlates of the subjective perception of oneself belonging to a particular gender. In most of these between-subject, group level studies, putative nuisance variables or confounds were regressed out from the data before proceeding to inference, and all of them used some form of permutation test to assess the significance of the results. In a recent review, \citet{Wang2020} described a permutation procedure for \textsc{cca} as ``a random shuffling of the rows (\ldots) of the two variable sets''.

Permutation tests are well known and widely used. Among their many benefits, these tests lead to valid inferences while requiring assumptions that are commonly satisfied in between-subject analyses, such that of exchangeability of observations under the null hypothesis. However, here we show that simple implementations of permutation inference for \textsc{cca} are inadequate on four different grounds. First, simple, uncorrected permutation p-values are not guaranteed to be monotonically related to the canonical correlations, leading to inadmissible results; for the same reason, multiple testing correction using false discovery rate is also inadmissible. Second, except for the highest canonical correlation, a simple one-step estimation of all others without considering the variability already explained by previous canonical variables in relation to each of them also leads to inflated per comparison error rates and thus, invalid results. Third, regressing out nuisance variables without consideration to the introduction of dependencies among observations caused by residualisation leads to an invalid test, with excess false positives. Fourth, multiple testing correction using the distribution of the maximum test statistic leads to conservative results, except for the highest canonical correlation.

In this paper we explain and discuss in detail each of these problems, and offer solutions that address each of them. In particular, we propose a stepwise estimation method for the canonical correlations and canonical variables that remains valid even when the number of variables is not the same for both sides of \textsc{cca}. We propose a method that transforms residualised data to a lower dimensional basis where exchangeability --- as required for the validity of permutation tests --- holds. We also propose that inference that considers multiple canonical correlations should use a closed testing procedure that is more powerful than the usual correction method used in permutation tests that use the distribution of the maximum statistic; the procedure also ensures a monotonic relationship between p-values and canonical correlations. Finally, we provide a complete, general algorithm for valid inferences for \textsc{cca}.

\section{Theory}
\label{sec:theory}

\subsection{Notation and general aspects}
\label{sec:theory:notation}

Thorough definition and derivation of \textsc{cca} can be found in many classical textbooks on multivariate analysis \citep[e.g.,][]{Kendall1975, Mardia1979, Brillinger1981, Muirhead1982, Seber1984, Krzanowski1988, Anderson2003}; the reader is referred to these for a comprehensive overview. Here we present concisely and only allude to the distinction between population ($\rho$) and sample ($r$) canonical correlations where strictly needed. Let $\mathbf{Y}_{N \times P}$ and $\mathbf{X}_{N \times Q}$ be each one a collection of, respectively, $P$ and $Q$ variables observed from $N$ subjects, $N \geqslant P+Q$. Without loss of generality, assume that $P \leqslant Q$, that the columns of $\mathbf{Y}$ and $\mathbf{X}$ are mean-centered, that these matrices are of full rank, and define $E(\mathbf{Y}'\mathbf{Y}) = \boldsymbol{\Sigma}_{\mathbf{YY}}$, $E(\mathbf{X}'\mathbf{X}) = \boldsymbol{\Sigma}_{\mathbf{XX}}$, and $E(\mathbf{Y}'\mathbf{X}) = \boldsymbol{\Sigma}_{\mathbf{YX}} = \boldsymbol{\Sigma}'_{\mathbf{XY}}$. The goal of \textsc{cca} is to identify \emph{canonical coefficients} or \emph{canonical weights} $\mathbf{A}_{P \times K} = [\mathbf{a}_{1}, \ldots, \mathbf{a}_{K}]$ and $\mathbf{B}_{Q \times K} = [\mathbf{b}_{1}, \ldots, \mathbf{b}_{K}]$, $K = \min(P,Q)$, such that the pairs $(\mathbf{u}_{k},\mathbf{v}_{k})$ of \emph{canonical variables}, defined as:

\begin{equation}
\begin{array}{lclcl}
\left[\mathbf{u}_{1}, \ldots, \mathbf{u}_{K}\right] &=& \mathbf{U}_{N \times K} &=& \mathbf{Y}\mathbf{A}\\
\left[\mathbf{v}_{1}, \ldots, \mathbf{v}_{K}\right] &=& \mathbf{V}_{N \times K} &=& \mathbf{X}\mathbf{B}
\end{array}
\label{eqn:UV}
\end{equation}

\noindent
have correlations $r_{k}$ that are maximal, under the constraint that $\mathbf{U}'\mathbf{U} = \mathbf{V}'\mathbf{V} = \mathbf{I}$. Estimation of $\mathbf{A}$ and $\mathbf{B}$ amounts to finding the $K$ solutions to:

\begin{equation}
\left[
\begin{array}{cc}
-r_{k}\boldsymbol{\Sigma}_{\mathbf{YY}} & \boldsymbol{\Sigma}_{\mathbf{YX}}\\
\boldsymbol{\Sigma}_{\mathbf{XY}}  & -r_{k}\boldsymbol{\Sigma}_{\mathbf{XX}}
\end{array}
\right] \left[
\begin{array}{c}
\mathbf{a}_{k} \\
\mathbf{b}_{k}
\end{array}
\right] = \mathbf{0}
\label{eqn:eigroots}
\end{equation}

\noindent
where the unknowns are $r_{k}$, $\mathbf{a}_{k}$ and $\mathbf{b}_{k}$; $r_{k}$ are the sample \emph{canonical correlations}, i.e., the correlations between the estimated canonical variables $\mathbf{u}_{k}$ and $\mathbf{v}_{k}$. The coefficients $\mathbf{a}_{k}$ are eigenvectors of $\boldsymbol{\Sigma}_{\mathbf{YY}}^{-1}\boldsymbol{\Sigma}_{\mathbf{YX}}\boldsymbol{\Sigma}_{\mathbf{XX}}^{-1}\boldsymbol{\Sigma}_{\mathbf{XY}}$, whereas $\mathbf{b}_{k}$ are eigenvectors of $\boldsymbol{\Sigma}_{\mathbf{XX}}^{-1}\boldsymbol{\Sigma}_{\mathbf{XY}}\boldsymbol{\Sigma}_{\mathbf{YY}}^{-1}\boldsymbol{\Sigma}_{\mathbf{YX}}$; the respective eigenvalues (for either $\mathbf{a}_{k}$ or $\mathbf{b}_{k}$, as these eigenvalues are the same) are $r_{k}^2$. For convenience, we call \emph{canonical component} the ensemble formed by the $k$-th canonical correlation, its corresponding pair of canonical variables, and associated pair of canonical coefficients; canonical variables may also be termed \emph{modes of variation}.

The typical method for estimation involves an iterative procedure that finds one $r_{k}$ and $\mathbf{a}_{k}$ at a time, with $\mathbf{b}_{k}$ computed as a function of these. However, the method proposed by \citet{Bjorck1973} is more concise and numerically more stable; it is described in the Appendix (Algorithm 3). The canonical correlations are then produced in descending order, $r_{1} \geqslant \ldots \geqslant r_{k} \geqslant \ldots \geqslant r_{K} \geqslant 0$; this positiveness of all canonical correlations is a consequence of these values being explicitly maximised during estimation; reversal of the sign of the coefficients $\mathbf{a}_k$ can always be accompanied by the reversal of the sign of the corresponding coefficients $\mathbf{b}_k$ in the other side (and of $\mathbf{u}_k$ and $\mathbf{v}_k$), to no net effect on $r_k$. That is, the signs of the canonical variables and coefficients are indeterminate, and any solution is arbitrary; nothing can be concluded about the specific direction of effects with \textsc{cca}.

\subsection{Parametric inference}
\label{sec:theory:parametric}

The distribution of the sample canonical correlations $r_k$ is intractable, even under assumptions of normality and independence among subjects, and is a function of the population correlations $\rho_k$ \citep{Constantine1963, James1964}. This difficulty led to the development of a rich asymptotic theory \citep{Fisher1939, Hsu1941, Lawley1959, Fujikosh1977, Glynn1978}. However, these approximations have been shown to be extremely sensitive to departures from normality, or require additional terms that further complicate their use \citep{Muirhead1980}; \citet{Brillinger1981} recommended resampling methods to estimate parameters used by normal approximations, which otherwise can be biased \citep{Anderson2003}. These difficulties pose challenges for inference. Even though some computationally efficient algorithms have been proposed \citep{Koev2006}, these tests continue to be rarely used.

Instead, a test based on whether a certain number of correlations are equal to zero has been proposed. The null hypothesis is $\mathcal{H}^0_{k} : \rho_{k} = \rho_{k+1} = \ldots = \rho_K = 0$, i.e., $\mathcal{H}^0_{k} :\wedge _{i=k}^{K} \rho_{i} = 0$, for $1 \leqslant k \leqslant K$, that is, the null is that $K-k+1$ population canonical correlations (the smaller ones) are zero \citep{Bartlett1938, Bartlett1947, Marriott1952, Lawley1959, Fujikoshi1974}, versus the alternative that at least one is not, i.e., $\mathcal{H}^1_{k} : \vee _{i=k}^{K} \rho_{i} > 0$. The test is based on the statistic proposed by \citet{Wilks1935}, as:

\begin{equation}
\lambda_{k} = -\left(N-C-\frac{P+Q+3}{2}\right)\ln\left(\prod_{i=k}^{K}(1-r_{i}^{2})\right)
\label{eqn:Wilks}
\end{equation}

\noindent
where the constant $C=0$ if there are no nuisance variables (Section \ref{sec:theory:nuisance}). Under the null hypothesis, $\lambda_{k}$ follows an approximate $\chi^2$ distribution with degrees of freedom $\nu = (P-k+1)(Q-k+1)$ if each of the columns of $\mathbf{Y}$ and $\mathbf{X}$ have values that are independent and identically distributed following a normal distribution \citep[but see][for a different expression]{Glynn1978}. Unfortunately, this test is sensitive to departures from normality, particularly in the presence of outliers, and its use has been questioned \citep{Seber1984, Harris1976}.

Another test statistic is based on \citet{Roy1953}\footnote{\citet{Roy1953} proposed two distinct but related test statistics; these are both known as ``Roy's largest root''. Here we use the one that is interpreted as a coefficient of determination, and not the other that is interpreted as an $F$-statistic. See \citet{Kuhfeld1986} for a complete discussion.}, and is simply:

\begin{equation}
\theta_{k} = r_{k}^{2}
\label{eqn:Roy}
\end{equation}

\noindent
The critical values for the corresponding parametric distribution at a given test level $\alpha$ can be found in the charts provided by \citet{Heck1960}, using as parameters $s=\min(P,Q)-k+1$, $m=(\left|P-Q\right|-1)/2$, and $n=(N-C-P-Q-2)/2$ \citep{Lee1978}, where the constant $C=0$ if there are no nuisance variables (Section \ref{sec:theory:nuisance}), or in tables provided by \citet{Kres1975}; more recent approximations for normally distributed data can be found in \citet{Chiani2016} and \citet{Johnstone2017}. Some approximations, however, are considered conservative \citep{Harris1976, Harris2013}. Note that, while \citet{Roy1953} proposed the use of the largest value as test statistic, which would then be $r_{k=1}^{2}$, any given null $\mathcal{H}^0_{k}$ at position $k$ must have already considered the previous canonical components, from $1$ until $k-1$, such that the maximum statistic, after the previous canonical correlations have been removed from the model, is always the current one. A similar reasoning holds for the smallest canonical correlations in the test proposed by \citet{Wilks1935}. This feature is exploited in the stepwise rejective procedure proposed in Section \ref{sec:theory:multiplicity}.

\subsection{Permutation inference}
\label{sec:theory:permutation}

The above problems can be eschewed with the use of resampling methods, such as permutation. An intuitive (but inadequate) permutation test for \textsc{cca} could be constructed by randomly permuting the rows of $\mathbf{Y}$ or $\mathbf{X}$. For each shuffling of the data, indicated by $j = \left\{1, \ldots, J\right\}$, a new set of canonical correlations $(r_{k})^{*}_{j}$ and associated statistics $(\lambda_{k})^{*}_{j}$ would be computed. A p-value would be obtained as $p_{k} = \frac{1}{J} \sum_{j=1}^J I\left[(\lambda_{k})_{1} \geqslant (\lambda_{k})^{*}_{j}\right]$, where $I\left[\cdot\right]$ is the indicator (Kronecker) function, which evaluates as 1 if the condition inside the brackets is true, or 0 otherwise, and the index $j=1$ corresponds to the unpermuted data (i.e., no permutation, with the data in their original ordering).

Such a na\"{i}ve procedure, however, would ignore the fact that, this resampling scheme matches the first null hypothesis $\mathcal{H}^0_{1}$, but not the subsequent ones. For a given canonical correlation at position $k$ being tested, $k>1$, one must generate a permutation that satisfy the corresponding null $\mathcal{H}^0_{k}$, but not necessarily $\{\mathcal{H}^0_{1}, \ldots, \mathcal{H}^0_{k-1}\}$. Otherwise, latent effects represented by the corresponding earlier canonical variables $[\mathbf{u}_{1}, \ldots, \mathbf{u}_{k-1}]$ and $[\mathbf{v}_{1}, \ldots, \mathbf{v}_{k-1}]$ would, in the procedure above, remain in the $\mathbf{Y}$ and $\mathbf{X}$ at the time these are permuted. However, the variance associated with these earlier canonical variables would have already been explained through the rejection of their respective null hypothesis up to $\mathcal{H}^{0}_{k-1}$. This variance is now a nuisance for the positions from $k$ (inclusive) onward. It contains information that are not pertinent to position $k$ and subsequent ones, and that therefore should not be used to build the null distribution, i.e., variance should not be re-used in the shufflings that lead to a given $(r_{k})^{*}_{j}$ or subsequent correlations.

Fortunately, \textsc{cca} is invariant to linear transformations that mix the variables in $\mathbf{Y}$ or in $\mathbf{X}$. Since the canonical variables are themselves linear transformations of these input variables (Equation \ref{eqn:UV}), a second \textsc{cca} using $\mathbf{U}$ and $\mathbf{V}$ in place of the initial $\mathbf{Y}$ and $\mathbf{X}$ leads to the same solutions. Yet, unless $P = Q$, $\mathbf{V}$ will not span the same space as $\mathbf{X}$. In principle, this would be inconsequential as far as the canonical variables are concerned. However, ignoring the variability in $\mathbf{X}$ not contained in $\mathbf{V}$ would again affect the p-values should $\mathbf{U}$ and $\mathbf{V}$ be used in a permutation test, as the permuted data would not be representative of the original (unpermuted) that led to these initial canonical variables. To mitigate the problem, include into the matrix of canonical coefficients their orthogonal complement, i.e., compute $\mathbf{\tilde{V}} = \mathbf{X}\left[ \mathbf{B}, \text{null}\left(\mathbf{B}'\right)\right]$, then use $\mathbf{\tilde{V}}$ instead of $\mathbf{V}$ as a replacement for $\mathbf{X}$. In this paper we adopted the convention that $P \leqslant Q$, but the same procedure works in reverse and, algorithmically, it might as well be simpler to compute also a $\mathbf{\tilde{U}} = \mathbf{Y}\left[ \mathbf{A}, \text{null}\left(\mathbf{A}'\right)\right]$ and use it instead of $\mathbf{U}$ as a replacement for $\mathbf{Y}$. If $P \leqslant Q$, then $\mathbf{\tilde{U}} = \mathbf{U}$.

While these transformations do not change in any way the canonical components, they allow the construction of an improved algorithm that addresses the issue of variability already explained by canonical variables of lower rank (i.e., the ones with order indices smaller than that of a given one). It consists of running an initial \textsc{cca} using $\mathbf{Y}$ and $\mathbf{X}$ to obtain $\mathbf{\tilde{U}}$ and $\mathbf{\tilde{V}}$, then subject these to a second \textsc{cca} and permutation testing while, crucially, at each permutation, iteratively repeating \textsc{cca} $K$ times, each using not the whole $\mathbf{\tilde{U}}$ and $\mathbf{\tilde{V}}$, but only from the $k$-th component onwards, i.e., $[\mathbf{\tilde{u}}_{k}, \ldots, \mathbf{\mathbf{\tilde{u}}}_{P}]$ and $[\mathbf{\tilde{v}}_{k}, \ldots, \mathbf{\tilde{v}}_{Q}]$ for the test about the $k$-th canonical correlation. Of note, a test level $\alpha$ does not need to be specified at the time in which the above iterative (stepwise) procedure is performed; instead, and in combination with the multiple testing procedure described below, adjusted p-values are are computed, which then are used to accept or reject the null for the $k$-th correlation. Algorithm 1 (Section \ref{sec:algorithm}) shows the procedure in detail (the algorithm contains other details that are discussed in the next sections).

A number of further aspects need be considered in permutation tests: the number of possible reorderings of the data, the need for permutations that break the association between the variables being tested, the random selection of permutations from the permutation set when not all possible permutations can be used, the choice of the test statistic, how to correct for the multiplicity of tests, the number of permutations to allow narrow confidence intervals around p-values, among others. These topics have been discussed in \citet{Winkler2014, Winkler2016} and references therein and will not be repeated here. However, for \textsc{cca}, some aspects deserve special treatment and are considered below.

\subsection{Choice of the statistic}
\label{sec:theory:statistic}

Asymptotically, using Wilks' statistic $\lambda_{k}$ or Roy's $\theta_{k}$ are expected to lead to the same conclusion since all correlations are sorted in descending order: if $r_k=0$, then all subsequent ones must be zero; likewise, if $r_k>0$, then clearly at least one correlation between $k$ and $K$ is larger than zero, which has to include $r_k$ itself. Moreover, permutation under the null is justifiable in the complete absence of association between the two sets, which implies that, under the null $\mathcal{H}^0_k$, all correlations $r_{k}, r_{k+1}, \ldots, r_{K}$ are equal to zero. With finite data, however, one statistic can be more powerful than the other in different settings; their relative performance is studied in Sections \ref{sec:evaluation} and~\ref{sec:results}.

Computationally, Wilks' requires more operations to be performed compared to Roy's statistic. Since the relationship between $r_k$ and $\theta_k$ is monotonic, the two are permutationally equivalent, and using $r_k$ alone is sufficient, which makes Roy's the absolute fastest. However, even for Wilks', the amount of computation required is negligible compared to the overall number of operations needed for estimation of the canonical coefficients, such that in practice, the choice between the two should be in terms of power.

In either case, while inference refers to the respective null hypothesis at position $k$, it is not to be understood as inference on the index $k$. Rather, the null is merely conditional on the nulls for all previous correlations from $1$ to $k-1$ having been rejected. Rejecting the null implies that the correlation observed at position $k$ is too high under the null hypothesis of no association between the two variable sets after all previous (from $1$ to $k-1$) canonical variables have been sequentially removed, as described in Section \ref{sec:theory:permutation}. In Algorithm 1 (Section \ref{sec:algorithm}) this is done in line 29, that uses as inputs to \textsc{cca} the precomputed canonical variables only from position $k$ onwards, as opposed to all of them.

\subsection{Multiplicity of tests}
\label{sec:theory:multiplicity}

For either of these two test statistics, the ordering of the canonical correlations from larger (farther from zero) to smaller (closer to zero) imply that rejection of the null hypothesis at each $k$ must happen sequentially, starting from $k=1$, using the respective test statistic and associated p-value until the null $\mathcal{H}^{0}_{k}$, for some $k=\{1, \ldots, K\}$, is not rejected at a predefined test level $\alpha$. Then, at that position $k$, the procedure stops, and the null is retained from that position (inclusive) onward until the final index $K$.

The ordering of the canonical correlations brings additional consequences. First, because rejection of $\mathcal{H}^{0}_{k}$ implies rejection of all joint (intersection) hypotheses that include $\mathcal{H}^{0}_{k}$, that is $\mathcal{H}^{0}_{1}, \ldots, \mathcal{H}^{0}_{k-1}$, such sequentially rejective procedure is also a closed testing procedure (\textsc{ctp}), which controls the amount of any type \textsc{i} error across all tests, i.e., the \emph{familywise error rate} (\textsc{fwer}) in the strong sense \citep{Marcus1976, Hochberg1987}. Thus, there is no need for further correction for multiple testing. Another way of stating the same is that the test for a given $r_k$, $k>1$, has been ``protected'' by the test at the position $k=1$ at the level $\alpha$. Adjusted p-values (in the \textsc{fwer} sense) can be computed as $[p_{k}]_{\textsc{fwer}}=\max(p_{1},\ldots,p_{k})$, that is, the \textsc{fwer}-adjusted p-value for the canonical component $k$ is the cumulative maximum p-value up to position $k$. Such adjusted p-values can be considered significant if their value is below the desired test level $\alpha$.

The second consequence is that $\textsc{fwer}$ adjustment of p-values using the distribution of the maximum statistic (not to be confused with the cumulative maximum described in the above paragraph) will be conservative for all canonical components except the first. The reason is that the maximum statistic is always the distribution of the first, which is stochastically dominant over all others. The distribution of the maximum is usually sought as a shortcut to a \textsc{ctp} when the condition of subset pivotality holds \citep{Westfall1993}, as that reduces the computational burden from $2^K$ tests to only $K$ tests. Interestingly, the ordering of the canonical correlations from largest to smallest leads to a \textsc{ctp} that does not use the distribution of the maximum, and yet requires only $K$ tests, regardless of whether subset pivotality holds.

A third consequence is that using permutation p-values outside the above sequentially rejective procedure that controls \textsc{fwer} is not appropriate since these simple, uncorrected p-values are not guaranteed to be monotonically related to the canonical correlations $r_{k}$. Attempts to use these uncorrected p-values outside a \textsc{ctp} would lead to paradoxical results whereby higher, stronger canonical correlations might not be considered significant, yet later ones, smaller, weaker, could be so; that is, it would make the test \emph{inadmissible} \citep[p.\ 232]{Lehmann2005}. For the same reason, such simple p-values should not be subjected to correction using false discovery rate \citep[\textsc{fdr};][]{Benjamini1995}, because the ordering of p-values for \textsc{fdr}, from smallest to largest, is not guaranteed to match the ordering of the canonical correlations, leading similarly to an inadmissible test.

\subsection{Nuisance variables}
\label{sec:theory:nuisance}

Few authors discussed nuisance variables or confounds in canonical correlation analysis, e.g., \citet{Roy1957, Rao1969, Timm1976, Lee1978, Sato2010}. Let the $\mathbf{Z}$ be a $N \times R$ matrix of nuisance variables, including an intercept. \emph{Partial} \textsc{cca} consists of considering $\mathbf{Z}$ nuisance for both $\mathbf{Y}$ and $\mathbf{X}$. This is distinct from \emph{part} \textsc{cca}, which consists of considering $\mathbf{Z}$ a nuisance for either $\mathbf{Y}$ or $\mathbf{X}$, but not both. Finally, \emph{bipartial} \textsc{cca} consists of considering $\mathbf{Z}$ a nuisance for $\mathbf{Y}$, while considering another set of variables $\mathbf{W}$, of size $N \times S$, a nuisance for $\mathbf{X}$. In all three cases, such nuisance variables can be regressed out from the respective set of variables of interest, then the respective residuals subjected to \textsc{cca} (Table \ref{tab:nuisance}). In the parametric case, inference can proceed using the distribution of $\lambda_k$ or $\theta_{k}$ (Equations \ref{eqn:Wilks} and \ref{eqn:Roy}) using $C=R$ for partial or part, and $C=\max(R,S)$ for bipartial \textsc{cca} \citep{Timm1976, Lee1978}.

\begin{table}[!t]
\caption{Taxonomy of canonical correlation analysis with respect to nuisance variables. In all cases, the aim is to find linear combinations of variables in the left and in the right sets, such that each combination from one set is maximally correlated with a corresponding combination from the other, but uncorrelated with all other combinations from either set.}
\begin{center}
\begin{tabular}{@{}m{77mm}<{\raggedright}m{20mm}<{\raggedright}m{20mm}<{\raggedright}@{}}
\toprule
Name & Left set & Right set\\
\midrule
\textsc{cca} (``full'', no nuisance) & $\mathbf{Y}$                        & $\mathbf{X}$\\
Partial \textsc{cca}                 & $\mathbf{R}_{\mathbf{Z}}\mathbf{Y}$ & $\mathbf{R}_{\mathbf{Z}}\mathbf{X}$\\
Part \textsc{cca}                    & $\mathbf{R}_{\mathbf{Z}}\mathbf{Y}$ & $\mathbf{X}$\\
Bipartial \textsc{cca}               & $\mathbf{R}_{\mathbf{Z}}\mathbf{Y}$ & $\mathbf{R}_{\mathbf{W}}\mathbf{X}$\\
\bottomrule
\end{tabular}
\end{center}
{\footnotesize
$\mathbf{R}_{\mathbf{Z}}$ is a residual forming matrix that considers the nuisance variables in $\mathbf{Z}$, and is computed as $\mathbf{I}_{N \times N}-\mathbf{Z}\mathbf{Z}^{+}$, where the symbol $^{+}$ represents a pseudo-inverse. $\mathbf{R}_{\mathbf{W}}$ is computed similarly, considering the nuisance variables in $\mathbf{W}$. The choice which set is on left or right side is arbitrary.
\par}
\label{tab:nuisance}
\end{table}

Permutation inference, however, requires further considerations, otherwise, as shown in Section \ref{sec:results}, results will be invalid. Consider first the case without nuisance variables. Let $\mathbf{M}=[\mathbf{Y}, \mathbf{X}]_{N \times (P+Q)}$ be the horizontal concatenation of the two sets of variables whose association is being investigated. Both $\mathbf{Y}$ and $\mathbf{X}$ occupy an $N$-dimensional space and, therefore, so does $\mathbf{M}$. A random permutation of the rows of either of the two sets of variables will not affect their dimensionalities. For example $\mathbf{M}^{*}=[\mathbf{P}\mathbf{Y}, \mathbf{X}]$ continues to occupy the same $N$-dimensional space as $\mathbf{M}$.

However, residualisation changes this scenario. Let $\mathbf{R}_{\mathbf{Z}} = \mathbf{I} - \mathbf{Z}\mathbf{Z}^{+}$ be the residual forming matrix associated with the nuisance variables $\mathbf{Z}$, with the symbol $^{+}$ representing the Moore--Penrose pseudo-inverse. $\mathbf{R}_{\mathbf{Z}}$ has the following interesting properties: $\mathbf{R}_{\mathbf{Z}}=\mathbf{R}_{\mathbf{Z}}'$ (symmetry) and $\mathbf{R}_{\mathbf{Z}}\mathbf{R}_{\mathbf{Z}}=\mathbf{R}_{\mathbf{Z}}$ (idempotency), both of which will be exploited later. In partial \textsc{cca}, $\mathbf{Z}$ can be regressed out from $\mathbf{Y}$ and $\mathbf{X}$ by computing $\mathbf{\tilde{Y}} = \mathbf{R}_{\mathbf{Z}}\mathbf{Y}$ and $\mathbf{\tilde{X}} = \mathbf{R}_{\mathbf{Z}}\mathbf{X}$. Let $\mathbf{\tilde{M}}=[\mathbf{\tilde{Y}}, \mathbf{\tilde{X}}]$ be the concatenation of the residualised sets $\mathbf{Y}$ and $\mathbf{X}$ with respect to $\mathbf{Z}$. While $\mathbf{Y}$ occupies an $N$-dimensional space, $\mathbf{\tilde{Y}}$ occupies a smaller one; its dimensions are, at most, of a size given by the rank of $\mathbf{R}_{\mathbf{Z}}$, which is $N-R$ assuming $\mathbf{Y}$ and $\mathbf{Z}$ are of full rank. The same holds for $\mathbf{X}$ and $\mathbf{\tilde{X}}$ and, therefore, for $\mathbf{M}$ and $\mathbf{\tilde{M}}$.

Permutation affects these relations: while $\mathbf{M}^{*}$ still occupies a space of $N$ dimensions as the unpermuted $\mathbf{M}$, $\mathbf{\tilde{M}}^{*}$, differently than $\mathbf{\tilde{M}}$, may now occupy a space with dimensions anywhere between $N-R$ and $N$, depending on a given random permutation. With fewer effective observations determined by this lower space after residualisation, and the same number of variables, the sample canonical correlations in the unpermuted case are stochastically larger than in the permuted, which in turn leads to an excess of spuriously small p-values. For not occupying the same space as the original, the permuted data are no longer a similar realisation of the unpermuted, thus violating exchangeability, and specifically causing the distribution of the test statistics to be unduly shifted to the left.

Here the following solution is proposed: using the results from \citet{Huh2001}, let $\mathbf{Q}_{\mathbf{Z}}$ be a $N \times N'$ semi-orthogonal basis \citep[p.\ 84]{Abadir2005} for the column space of $\mathbf{R}_{\mathbf{Z}}$ constructed via, e.g., spectral or Schur decomposition, such that $\mathbf{Q}_{\mathbf{Z}}\mathbf{Q}'_{\mathbf{Z}}=\mathbf{R}_{\mathbf{Z}}$, $\mathbf{Q}'_{\mathbf{Z}}\mathbf{Q}_{\mathbf{Z}}=\mathbf{I}_{N' \times N'}$, where $N'=N-R$, and $\mathbf{Q}'_\mathbf{Z}=\mathbf{Q}_{\mathbf{Z}}^{+}$. Then $\textsc{cca}$ on $\mathbf{\tilde{\tilde{Y}}}=\mathbf{Q}_{\mathbf{Z}}'\mathbf{Y}$ and $\mathbf{\tilde{\tilde{X}}}=\mathbf{Q}_{\mathbf{Z}}'\mathbf{X}$ lead to the same solutions as on $\mathbf{\tilde{Y}}=\mathbf{R}_{\mathbf{Z}}\mathbf{Y}$ and $\mathbf{\tilde{X}}=\mathbf{R}_{\mathbf{Z}}\mathbf{X}$. The reason is that, from Section \ref{sec:theory:notation}, $E((\mathbf{R}_{\mathbf{Z}}\mathbf{Y})'(\mathbf{R}_{\mathbf{Z}}\mathbf{Y})) = E(\mathbf{Y}'\mathbf{R}_{\mathbf{Z}}'\mathbf{R}_{\mathbf{Z}}\mathbf{Y}) = E(\mathbf{Y'R_ZY})$, which is the same as $E((\mathbf{Q}_{\mathbf{Z}}'\mathbf{Y})'(\mathbf{Q}_\mathbf{Z}'\mathbf{Y})) = E(\mathbf{Y}'\mathbf{Q}_{\mathbf{Z}}\mathbf{Q}_{\mathbf{Z}}'\mathbf{Y}) = E(\mathbf{Y}'\mathbf{R}_{\mathbf{Z}}\mathbf{Y})$, since, as discussed earlier, $\mathbf{R}_{\mathbf{Z}}$ is symmetric and idempotent, and $\mathbf{Q}_{\mathbf{Z}}\mathbf{Q}'_{\mathbf{Z}}=\mathbf{R}_{\mathbf{Z}}$. In a similar manner, $E((\mathbf{Q}_{\mathbf{Z}}'\mathbf{X})'(\mathbf{Q}_{\mathbf{Z}}'\mathbf{X})) = E(\mathbf{X}'\mathbf{R}_{\mathbf{Z}}\mathbf{X})$, and likewise, $E((\mathbf{Q}_{\mathbf{Z}}'\mathbf{X})'(\mathbf{Q}_{\mathbf{Z}}'\mathbf{Y})) = E(\mathbf{X}'\mathbf{R}_{\mathbf{Z}}\mathbf{Y})$. While pre-multi\-plication by $\mathbf{Q}_{\mathbf{Z}}'$ does not affect the \textsc{cca} results\footnote{As originally proposed, in the context of the general linear model (\textsc{glm}), \citet{Huh2001} use $\mathbf{\tilde{\tilde{Y}}}=\mathbf{Q}_{\mathbf{Z}}'\mathbf{R}_{\mathbf{Z}}\mathbf{Y}$ and $\mathbf{\tilde{\tilde{X}}}=\mathbf{Q}_{\mathbf{Z}}'\mathbf{R}_{\mathbf{Z}}\mathbf{X}$. These are equivalent to simply $\mathbf{\tilde{\tilde{Y}}}=\mathbf{Q}_{\mathbf{Z}}'\mathbf{Y}$ and $\mathbf{\tilde{\tilde{X}}}=\mathbf{Q}_{\mathbf{Z}}'\mathbf{X}$ as proposed here: since $\mathbf{R}_{\mathbf{Z}} = \mathbf{Q}_{\mathbf{Z}}\mathbf{Q}'_{\mathbf{Z}}$, $\mathbf{Q}_{\mathbf{Z}}'\mathbf{R}_{\mathbf{Z}} = \mathbf{Q}_{\mathbf{Z}}'\mathbf{Q}_{\mathbf{Z}}\mathbf{Q}_{\mathbf{Z}}' = \mathbf{I}_{N' \times N'}\mathbf{Q}_{\mathbf{Z}}'=\mathbf{Q}_{\mathbf{Z}}'$. This simplification holds true also for the \textsc{glm} (not discussed in this article).}, it changes the dependence structure among the rows of the data: $\mathbf{\tilde{\tilde{M}}}=[\mathbf{\tilde{\tilde{Y}}}, \mathbf{\tilde{\tilde{X}}}]$ occupies an $N'$-dimensional space, and so does $\mathbf{\tilde{\tilde{M}}}^{*}=[\mathbf{P}\mathbf{\tilde{\tilde{Y}}}, \mathbf{\tilde{\tilde{X}}}]$, for a permutation matrix $\mathbf{P}$ of size $N' \times N'$, such that exchangeability holds, thus allowing a valid permutation test. 

The treatment of partial \textsc{cca}, as described above, can be seen as a particular case of bipartial \textsc{cca} in which $\mathbf{W}=\mathbf{Z}$, that is, the set of nuisance variables in both sides is the same. Of course, for bipartial \textsc{cca} proper, this equality does not necessarily hold, and the two sets may be different in different ways: $\mathbf{Z}$ may be entirely orthogonal to $\mathbf{W}$, or some or all variables from one set may be fully represented in the other, either directly (e.g., some of the variables present in both sets), or as linear combinations of one set in the other, or it may be that these two sets are simply not orthogonal. The direct strategy of computing $\mathbf{R}_\mathbf{W} = \mathbf{I}-\mathbf{WW}^+$ and its respective semi-orthogonal matrix $\mathbf{Q}_\mathbf{W}$ leads to difficulties because, unless $R = S$, the products $\mathbf{\tilde{\tilde{Y}}}=\mathbf{Q}_{\mathbf{Z}}'\mathbf{Y}$ and $\mathbf{\tilde{\tilde{X}}}=\mathbf{Q}_{\mathbf{W}}'\mathbf{X}$ will not have the same number of rows: $\mathbf{\tilde{\tilde{Y}}}$ has $N'=N-R$, whereas $\mathbf{\tilde{\tilde{X}}}$ has $N''=N-S$ rows, thus preventing the computation of \textsc{cca}.

A more general solution, that accommodates bipartial and, therefore, is a generalisation for all cases of nuisance variables in \textsc{cca}, consists of randomly permuting rows of $\mathbf{\tilde{\tilde{Y}}}$ and/or $\mathbf{\tilde{\tilde{X}}}$ using, respectively, permutation matrices $\mathbf{P}_{\mathbf{Y}}$ and $\mathbf{P}_{\mathbf{X}}$ of respective sizes $N'$ and $N''$, therefore permuting in the lower dimensional space where $\mathbf{\tilde{\tilde{Y}}}$ and $\mathbf{\tilde{\tilde{X}}}$ are exchangeable, then, crucially, reestablishing the original number $N$ of rows using the property that the transpose of a semi-orthogonal matrix is the same as its inverse ($\mathbf{Q'}=\mathbf{Q^+}$), to only then perform \textsc{cca}. Therefore, \textsc{cca} is computed using $\mathbf{Q}_{\mathbf{Z}}\mathbf{P}_{\mathbf{Y}}\mathbf{Q}_{\mathbf{Z}}'\mathbf{Y}$ and $\mathbf{Q}_{\mathbf{W}}\mathbf{P}_{\mathbf{X}}\mathbf{Q}_{\mathbf{W}}'\mathbf{X}$. Left and right sides will continue to have rank $N'$ and $N''$ respectively, will have already been permuted, and will both have $N$ rows. The procedure is fully symmetric in that, when the permutation matrices $\mathbf{P}_{\mathbf{Y}}$ and $\mathbf{P}_{\mathbf{X}}$ are both identity matrices (of sizes $N'$ and $N''$, respectively), which is equivalent to no permutation, the expressions for each side reduce to the residualised data $\mathbf{R}_{\mathbf{Z}}\mathbf{Y}$ and $\mathbf{R}_{\mathbf{W}}\mathbf{X}$. The concatenation $[\mathbf{Q}_{\mathbf{Z}}\mathbf{P}_{\mathbf{Y}}\mathbf{Q}_{\mathbf{Z}}'\mathbf{Y}, \mathbf{Q}_{\mathbf{W}}\mathbf{P}_{\mathbf{X}}\mathbf{Q}_{\mathbf{W}}'\mathbf{X}]$ has the same rank as that of  $[\mathbf{R}_{\mathbf{Z}}\mathbf{Y}, \mathbf{R}_{\mathbf{W}}\mathbf{X}]$, thus addressing the above problem of the unpermuted test statistic having a different and stochastically dominant distribution over that of the permuted data. Table \ref{tab:solution} summarises the proposed solution for all cases, including part \textsc{cca}.

\begin{table}[!t]
\caption{Proposed permutation method for the various cases of \textsc{cca}, with respect to nuisance variables.}
\begin{center}
\begin{tabular}{@{}m{50mm}<{\raggedright}m{34mm}<{\raggedright}m{34mm}<{\raggedright}@{}}
\toprule
Name & Left set & Right set\\
\midrule
\textsc{cca} (``full'', no nuisance) & 
$\mathbf{P}_{\mathbf{Y}}\mathbf{Y}$ & 
$\mathbf{X}$\\
Partial \textsc{cca}                 & 
$\mathbf{P}_{\mathbf{Y}}\mathbf{Q}_{\mathbf{Z}}'\mathbf{Y}$ &
$\mathbf{Q}_{\mathbf{Z}}'\mathbf{X}$\\
Part \textsc{cca}                    & 
$\mathbf{Q}_{\mathbf{Z}}\mathbf{P}_{\mathbf{Y}}\mathbf{Q}_{\mathbf{Z}}'\mathbf{Y}$ & 
$\mathbf{P}_{\mathbf{X}}\mathbf{X}$\\
Bipartial \textsc{cca}               & 
$\mathbf{Q}_{\mathbf{Z}}\mathbf{P}_{\mathbf{Y}}\mathbf{Q}_{\mathbf{Z}}'\mathbf{Y}$ &
$\mathbf{Q}_{\mathbf{W}}\mathbf{P}_{\mathbf{X}}\mathbf{Q}_{\mathbf{W}}'\mathbf{X}$\\
\bottomrule
\end{tabular}
\end{center}
{\footnotesize
$\mathbf{Q}_{\mathbf{Z}}$ is a $N \times N'$ semi-orthogonal basis for the column space of $\mathbf{R}_{\mathbf{Z}}$, such that $\mathbf{Q}_{\mathbf{Z}}\mathbf{Q}'_{\mathbf{Z}}=\mathbf{R}_{\mathbf{Z}}$, $\mathbf{Q}'_{\mathbf{Z}}\mathbf{Q}_{\mathbf{Z}}=\mathbf{I}_{N' \times N'}$, where $N'=N-R$, and $\mathbf{Q}'_\mathbf{Z}=\mathbf{Q}_{\mathbf{Z}}^{+}$. $\mathbf{Q}_{\mathbf{W}}$ is a $N \times N''$ similarly defined matrix for the column space of $\mathbf{R}_{\mathbf{W}}$, $N'' = N-S$. The bipartial \textsc{cca} case generalizes all others: for ``full'' \textsc{cca}, $\mathbf{R}_{\mathbf{W}}=\mathbf{R}_{\mathbf{Z}}=\mathbf{I}_{N \times N}$, and so, $\mathbf{Q}_{\mathbf{W}}=\mathbf{Q}_{\mathbf{Z}}=\mathbf{I}_{N \times N}$; for partial \textsc{cca}, $\mathbf{W}=\mathbf{Z}$; for part \textsc{cca} $\mathbf{R}_{\mathbf{W}}=\mathbf{I}_{N \times N}$, and so, $\mathbf{Q}_{\mathbf{W}}=\mathbf{I}_{N \times N}$. For full and partial, pre-multiplication by $\mathbf{Q}_{\mathbf{Z}}$ can be omitted since $\mathbf{Q}_{\mathbf{Z}}'\mathbf{Q}_{\mathbf{Z}}=\mathbf{I}_{N \times N}$, such that results do not change. Once these simplifications are considered, the general bipartial \textsc{cca} case reduces to the other three as shown in the Table. Full and partial have matching number of rows in both sides, such that only one side needs be permuted; part and bipartial, however, have at the time of the permutation a different number of rows in each side, such that both can be permuted separately through the use of suitably sized permutation matrices $\mathbf{P}_{\mathbf{Y}}$ and $\mathbf{P}_{\mathbf{X}}$; $\mathbf{P}_{\mathbf{Y}}$ is size $N \times N$ for full \textsc{cca}, and $N' \times N'$ for the three other cases; $\mathbf{P}_{\mathbf{X}}$ is size $N \times N$ for full and for part \textsc{cca}, and  $N'' \times N''$ for the two other cases.
\par}
\label{tab:solution}
\end{table}

\subsection{Restricted exchangeability}
\label{sec:restricted}

The above method uses the Huh--Jhun semi-orthogonal matrix applied to \textsc{cca} and leads to a valid permutation test provided that there are no dependencies among the rows of $\mathbf{M}$. That is, the method takes into account dependencies introduced by the regression of $\mathbf{Z}$ and/or $\mathbf{W}$ out from $\mathbf{Y}$ and/or $\mathbf{X}$, but not dependencies that might already exist in the data, and which generally preclude direct use of permutation tests. However, structured dependencies, such as those that may exist, for instance, in studies that involve repeated measurements, or for those in which participants do not constitute independent observations, e.g., sib-pairs, as in the Human Connectome Project \citep[\textsc{hcp};][]{VanEssen2012}, can be treated by allowing only those permutations that respect such dependency structure \citep{Winkler2015}. Unfortunately, the Huh--Jhun semi-orthogonal matrix does not respect such structure, blurring information from observations across blocks, and preventing the definition of a meaningful mapping from the $N$ original observations that define the block structure to the $N'$ or $N''$ observations that are ultimately permuted.\footnote{There is an exception: if $\mathbf{Z}$ has a block diagonal structure and the observations encompassed by such blocks coincide with the exchangeability blocks, then an algorithm that uses Huh--Jhun and block permutation can be constructed.}

\begin{figure}[!tb]
\centering
\hspace*{0mm}\includegraphics[width=100mm]{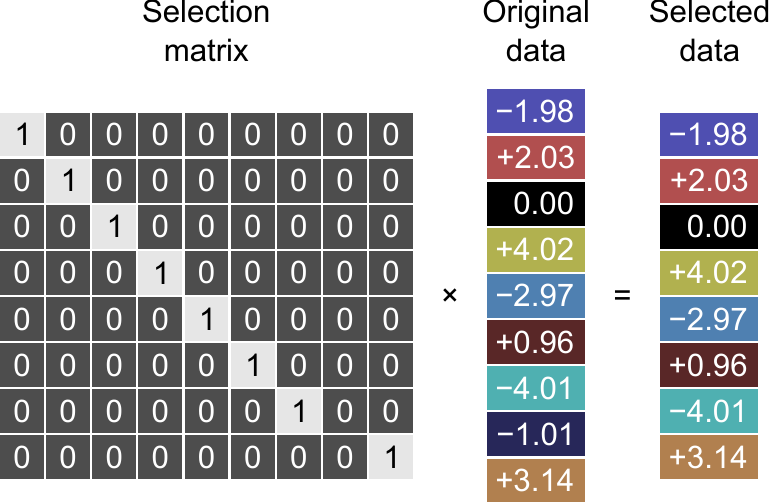}
\caption{A selection matrix is an identity matrix from which some specific rows have been removed. Pre-multiplication by a selection matrix deletes specific rows (those that correspond to columns that are all zero in the selection matrix).}
\label{fig:selection}
\end{figure}

Such mapping, whereby each one of the $N'$ and $N''$ rows of, respectively, $\mathbf{\tilde{\tilde{Y}}}$ and $\mathbf{\tilde{\tilde{X}}}$ corresponds uniquely to one of the $N$ rows of the original data $\mathbf{Y}$ and $\mathbf{X}$, can be obtained using a different method, due to \citet{Theil1965, Theil1968}, and reviewed in detail by \citet{Magnus2005}. Consider first the case of $\mathbf{Z}$. In the Theil method, that here is adapted for \textsc{cca}, $\mathbf{Q}_{\mathbf{Z}} = \mathbf{R}_{\mathbf{Z}}\mathbf{S}'(\mathbf{S}\mathbf{R}_{\mathbf{Z}}\mathbf{S}')^{-1/2}$, where the exponent $-1/2$ represents the positive definite matrix square root, and $\mathbf{S}$ is a $\bar{N} \times N$ selection matrix, $\bar{N}=\min(N',N'')$, that is, an identity matrix from which some $\max(R,S)$ rows have been removed. Pre-multiplication of a matrix by a selection matrix deletes specific rows, i.e., the ones that correspond to columns that are all zero in the selection matrix (Figure \ref{fig:selection}). The $\mathbf{Q}_{\mathbf{Z}}'\mathbf{Y}$ thusly computed are the best linear unbiased residuals with scalar covariance (\textsc{blus}), in that they are unbiased estimates of $\mathbf{S}\boldsymbol{\epsilon}_{\mathbf{Y}}$, where $\boldsymbol{\epsilon}_{\mathbf{Y}}$ are the (unknown) true errors after the nuisance effects of $\mathbf{Z}$ have been removed from $\mathbf{Y}$; $\mathbf{S}'\boldsymbol{\epsilon}_{\mathbf{Y}}$ contains the variance of interest, which may be shared among linear combinations of variables in both sides; it is an estimate of that is subjected to \textsc{cca} and statistical testing. For partial \textsc{cca}, $\mathbf{Q}_{\mathbf{Z}}$ is the same for both sides; for bipartial \textsc{cca}, similar computations hold for the other side, i.e., $\mathbf{Q}_{\mathbf{W}} = \mathbf{R}_{\mathbf{W}}\mathbf{S}'(\mathbf{S}\mathbf{R}_{\mathbf{W}}\mathbf{S}')^{-1/2}$. Table \ref{tab:semiortho} summarises the two methods.

\begin{table}[!tb]
\caption{The semi-orthogonal matrix $\mathbf{Q}$ ($\mathbf{Q}_\mathbf{Z}$ and/or $\mathbf{Q}_\mathbf{W}$, subscripts dropped), discussed in Sections \ref{sec:theory:nuisance} and \ref{sec:restricted}, is not unique. Two principled methods to obtain it are below.}
\begin{center}
\begin{tabular}{@{}m{70mm}<{\raggedright}m{52mm}<{\raggedright}@{}}
\toprule
Method & Matrix \\
\midrule
\citet{Theil1965} & $\mathbf{Q}=\mathbf{R}\mathbf{S}'(\mathbf{S}\mathbf{R}\mathbf{S}')^{-1/2}$ \\
\citet{Huh2001}   & $\mathbf{Q}\mathbf{E}\mathbf{Q}'=\mathbf{R}$ (via \textsc{svd} or Schur) \\
\bottomrule
\end{tabular}
\end{center}
{\footnotesize
$\mathbf{R}$ is the residual-forming matrix ($\mathbf{R}_{\mathbf{Z}}$ or $\mathbf{R}_{\mathbf{W}}$, for the respective set of nuisance variables, subscript dropped); since $\mathbf{R}$ is idempotent, all its eigenvalues (the diagonal elements of $\mathbf{E}$) are equal to 0 or 1. In the Theil method, $\mathbf{S}$ is a $N' = N \times (N-R)$ (for $\mathbf{Z}$) or $N'' = N \times (N-S)$ (for $\mathbf{W}$) selection matrix; the matrix square root (in the exponent $-1/2$) is the positive definite solution. In the Huh--Jhun method, after Schur or \textsc{svd} factorisations of $\mathbf{R}$ are computed, the $R$ or $S$ columns of $\mathbf{Q}$ that have corresponding zero eigenvalues in the diagonal of $\mathbf{E}$ are removed, such that $\mathbf{Q}$ computed from the factosisation is reduced from size $N \times N$ to $N \times N'$ or to $N \times N''$. At the end of these computations (see Algorithm 2, ``semiortho'', in the Appendix), for both methods $\mathbf{Q}'\mathbf{Q}=\mathbf{I}$, $\mathbf{Q}\mathbf{Q}'=\mathbf{R}$, and $\mathbf{Q}'=\mathbf{Q}^{+}$. Both methods aim at obtaining residuals with a scalar covariance matrix $\sigma^2\mathbf{I}$. Theil explictly seeks \textsc{blus} residuals. However, strictly, $\mathbf{S}$ does not need to be a selection matrix: choose $\mathbf{S}$ to be $\mathbf{Q}\mathbf{R}$ (not to be confused with \textsc{qr} decomposition) using $\mathbf{Q}$ computed with the Huh--Jhun approach. Then, following \citet[Theorem 2, p.\ 42]{Magnus2005}, it can be shown that Huh--Jhun also provides \textsc{blus} residuals.
\par}
\label{tab:semiortho}
\end{table}

To construct a permutation procedure for \textsc{cca} that respects the block structure, the Theil method can be used to compute $\mathbf{Q}$ instead of the Huh--Jhun approach. Choose $\max(R,S)$ observations to be removed from both sides (for partial \textsc{cca}, $R=S$ since $\mathbf{W}=\mathbf{Z}$). Construct the selection matrix $\mathbf{S}$ of size $\bar{N} \times N$, define the exchangeability blocks based on $\bar{N}$ observations, compute $\mathbf{Q}_{\mathbf{Z}}$ and $\mathbf{Q}_{\mathbf{W}}$ using the same $\mathbf{S}$ for both (for part \textsc{cca}, use the same strategy as for bipartial, replacing $\mathbf{R}_{\mathbf{W}}$ for $\mathbf{I}$), residualise (in the \textsc{blus} sense) the input variables by computing $\mathbf{\tilde{\tilde{Y}}}$ and $\mathbf{\tilde{\tilde{X}}}$. These have the same number of rows, and the dependencies among these rows is the same for both sides; thus, only one side needs be subjected to random permutations that respect such existing dependencies. Optionally, after permutation, the number $N$ of observations may be reestablished by pre-multiplication by $\mathbf{Q}_{\mathbf{Z}}$ and $\mathbf{Q}_{\mathbf{W}}$. Finally, \textsc{cca} is performed, with observation to the aspects discussed in Sections \ref{sec:theory:permutation} and \ref{sec:theory:multiplicity}. A detailed algorithm is presented in Section \ref{sec:algorithm}.

It remains to be decided how to select the $\max(R,S)$ observations to be dropped. In principle, any set could be considered for removal, provided that the removed rows of $\mathbf{Z}$ or $\mathbf{W}$ form a full rank matrix. Some informed choices, however, could be more powerful than others. One of the conclusions from \citet{Winkler2015} is that the complexity of the dependence structure and the ensuing restrictions on exchangeability leads to reductions in power. Thus, natural candidates for removal are observations that, once removed, cause the overall dependence structure to be simpler. For example, it is sometimes the case that some observations are so uniquely related to all others that there are no other observations like them in the sample. These observations, therefore, cannot be permuted with any other, or perhaps with only a few. Their contribution to hypothesis testing in the permutation case is minimal, and their removal are less likely to affect a decision on rejection of the null hypothesis. Consider for example a design that has many monozygotic, dizygotic, and non-twin pairs of subjects, and that in the sample, there happens to be a single pair of half-siblings. It is well known that, for heritable traits, genetic resemblance depends on the kinship among individuals; half-siblings are expected to have a different degree of statistical dependency among each other compared to each one of the other types of sibships in this sample. Thus, in there being just one such pair, it would be reasonable to prioritise it for exclusion, while keeping others.

\subsection{General algorithm}
\label{sec:algorithm}

A set of steps for permutation inference for \textsc{cca} is described in Algorithm 1. In it, input variables $\mathbf{Y}$ and $\mathbf{X}$ will have been mean-centered before the algorithm begins, or an intercept will have been included as nuisance variable in both $\mathbf{Z}$ and $\mathbf{W}$. $\mathcal{P} = \left\{(\mathbf{P}_{\mathbf{Y}},\mathbf{P}_{\mathbf{X}})_{j}\right\}$ is a set containing pairs of permutation matrices indexed by $j=\left\{1, \ldots, J\right\}$. In this set, the first permutation is always ``no permutation'', i.e., $(\mathbf{P}_{\mathbf{Y}},\mathbf{P}_{\mathbf{X}})_{j=1} = (\mathbf{I}_{N' \times N'}, \mathbf{I}_{N'' \times N''})$, such that $(\lambda_{k})_{j=1}^{*}=\lambda_{k}$, for all $k$. For the cases in which only one side of \textsc{cca} needs be permuted (Table \ref{tab:solution}), or for the cases in which $R=S$, or when there are dependencies among the data such that the Theil method is used to construct $\mathbf{Q}$ (Table \ref{tab:semiortho}), then $(\mathbf{P}_{\mathbf{X}})_j$ can be set as $\mathbf{I}$ for all $j$. Details on how $\mathcal{P}$ is defined in observance to the null hypothesis and respecting structured dependencies among the data have been discussed in \citet{Winkler2014, Winkler2015}. In the algorithm, $P$ can be larger, equal, or smaller than $Q$. Optional input arguments are the matrices with nuisance variables $\mathbf{Z}$ and $\mathbf{W}$, and the selection matrix $\mathbf{S}$. If $\mathbf{Z}$ is supplied but not $\mathbf{W}$, then the algorithm performs part or partial \textsc{cca}, depending on the Boolean argument \textsc{partial}; if both $\mathbf{Z}$ and $\mathbf{W}$ are supplied, the algorithm performs bipartial $\textsc{cca}$; if neither is supplied, then ``full'' \textsc{cca} is performed. If $\mathbf{S}$ is supplied, then the \textsc{blus} residuals based on Theil are used; otherwise, Huh--Jhun residuals are used. For either of these two cases, the semi-orthogonal matrix $\mathbf{Q}$ is computed using a separate, ancillary function named ``semiortho'', described in the Appendix.

\begin{myalgorithm}
\singlespacing
\noindent
\caption{Permutation inference for \textsc{cca}.}
\HRule
\vspace{3mm}
{\small
\begin{algorithmic}[1]
\Require $\mathbf{Y}_{N \times P}, \mathbf{X}_{N \times Q}, \mathcal{P}$. \textbf{Optional:} $\mathbf{Z}_{N \times R}, \mathbf{W}_{N \times S}, \mathbf{S}, \textsc{partial}$.
\Comment{Inputs.}
\State $K \leftarrow \min (P,Q)$
\Comment{Number of canonical components.}

\If{$\text{exist}(\mathbf{Z})$}
\Comment{If left-side nuisance were defined.}
\State $\mathbf{R}_{\mathbf{Z}} \leftarrow \mathbf{I} - \mathbf{Z}\mathbf{Z}^{+}$
\Comment{Residual forming matrix due to $\mathbf{Z}$.}
\State $[\mathbf{Q}_{\mathbf{Z}}]_{N \times N'} \leftarrow \text{semiortho}(\mathbf{R}_{\mathbf{Z}},\mathbf{S})$
\Comment{$\mathbf{Q}_{\mathbf{Z}}$ via H--J/Theil; $N'\!=\!N\!-\!R$.}
\Else 
\State [$\mathbf{Q}_{\mathbf{Z}}]_{N \times N'} \leftarrow \mathbf{I}_{N \times N}$
\Comment{$\mathbf{Q}_{\mathbf{Z}}$ is identity; $N'\!=\!N$.}
\EndIf

\If{$\neg\ \text{exist}(\mathbf{W}) \wedge \textsc{partial}$}
\Comment{If $\mathbf{W}$ not given, and this is partial \textsc{cca}.}
\State $\mathbf{W} \leftarrow \mathbf{Z}$
\Comment{Re-use $\mathbf{Z}$ as $\mathbf{W}$.}
\EndIf
\Comment{(otherwise, it is bipartial or part \textsc{cca}.)}

\If{$\text{exist}(\mathbf{W})$}
\Comment{If right-side nuisance were defined.}
\State $\mathbf{R}_{\mathbf{W}} \leftarrow \mathbf{I} - \mathbf{W}\mathbf{W}^{+}$
\Comment{Residual forming matrix due to $\mathbf{W}$.}
\State $[\mathbf{Q}_{\mathbf{W}}]_{N \times N''} \leftarrow \text{semiortho}(\mathbf{R}_{\mathbf{W}},\mathbf{S})$
\Comment{$\mathbf{Q}_{\mathbf{W}}$ via H--J/Theil; $N''\!=\!N\!-\!S$.}
\Else 
\State [$\mathbf{Q}_{\mathbf{W}}]_{N \times N''} \leftarrow \mathbf{I}_{N \times N}$
\Comment{$\mathbf{Q}_{\mathbf{W}}$ is identity; $N''\!=\!N$.}
\EndIf

\State $\mathbf{Y} \leftarrow \mathbf{Q}_{\mathbf{Z}}'\mathbf{Y}$
\Comment{Residualised $\mathbf{Y}$, exchangeable.}
\State $\mathbf{X} \leftarrow \mathbf{Q}_{\mathbf{W}}'\mathbf{X}$
\Comment{Residualised $\mathbf{X}$, exchangeable.}

\State $\mathbf{A}_{P \times K}, \mathbf{B}_{Q \times K}, \_ \leftarrow \text{cca}\left(\mathbf{Y}, \mathbf{X}, R, S\right)$
\Comment{Initial \textsc{cca} (see Algorithm 3).}
\State $\mathbf{U}_{N \times P} \leftarrow \mathbf{Y}\left[\mathbf{A}, \text{null}\left(\mathbf{A}'\right)\right]_{P \times P}$
\Comment{Canonical variables from residualised $\mathbf{Y}$.}
\State $\mathbf{V}_{N \times Q} \leftarrow \mathbf{X}\left[\mathbf{B}, \text{null}\left(\mathbf{B}'\right)\right]_{Q \times Q}$
\Comment{Canonical variables from residualised $\mathbf{X}$.}
\For{$k \in \{1, \ldots, K\}$}
\Comment{For each component.}
\State $c_{k} \leftarrow 0$
\Comment{Initialise a counter.}
\EndFor

\ForAll{$(\mathbf{P}_{\mathbf{Y}},\mathbf{P}_{\mathbf{X}})_{j} \in \mathcal{P}$}
\Comment{For each permutation}.
\For{$k \in \{1, \ldots, K\}$}
\Comment{For each component.}
\State $[\mathbf{u}_{k}, \ldots, \mathbf{u}_{P}]^{*}_{j} \leftarrow \mathbf{Q}_{\mathbf{Z}}(\mathbf{P}_{\mathbf{Y}})_{j}[\mathbf{u}_{k}, \ldots, \mathbf{u}_{P}]$
\Comment{Permute left side.}
\State $[\mathbf{v}_{k}, \ldots, \mathbf{v}_{Q}]^{*}_{j} \leftarrow \mathbf{Q}_{\mathbf{W}}(\mathbf{P}_{\mathbf{X}})_{j}[\mathbf{v}_{k}, \ldots, \mathbf{v}_{Q}]$
\Comment{Permute right side.}
\State $\_, \_, [r_{(1)},\ldots]_{j}^{*} \leftarrow \text{cca}\left([\mathbf{u}_{k}, \ldots]^{*}_{j}, [\mathbf{v}_{k}, \ldots]^{*}_{j}, R, S\right)$
\Comment{Main \textsc{cca}.}
\State $(\lambda_{k})_{j}^{*} \leftarrow \sum_{i=1}^{K-k+1}\ln(1-(r_{(i)}^{2})_{j}^{*})$
\Comment{Wilks' test statistic.}
\If{$(\lambda_{k})_{j}^{*} \geqslant (\lambda_{k})_{j=1}^{*}$}
\Comment{If statistic after permutation is larger.}
\State $c_{k} \leftarrow c_{k}+1$ 
\Comment{Increment the counter.}
\EndIf
\EndFor
\EndFor

\For{$k \in \{1, \ldots, K\}$}
\Comment{For each canonical component.}
\State $[p_{k}]_{\text{unc}} \leftarrow c_{k}/J$
\Comment{Uncorrected p-value.}
\State $[p_{k}]_{\textsc{fwer}} \leftarrow \max([p_{1},\ldots,p_{k}]_{\text{unc}})$
\Comment{\textsc{fwer}-corrected p-value (closure).}
\EndFor
\State \Return $[{p_{1}, \ldots, p_{K}}]_{\textsc{fwer}}$
\Comment{Return the \textsc{fwer}-corrected p-values.}
\end{algorithmic}}
\noindent
\HRule\\[2mm]
\setstretch{\lspac}
\end{myalgorithm}

An initial \textsc{cca} using residualised data is done in line 19; this uses another ancillary function, named ``cca'', and also described in the Appendix; this function returns three results: the canonical coefficients $\mathbf{A}$ and $\mathbf{B}$, and the canonical correlations $r_k$. The canonical coefficients are used to compute the canonical variables $\mathbf{U}$ and $\mathbf{V}$, augmented by their orthogonal complement needed to ensure that they span the same space as the variables subjected to this initial \textsc{cca}; the canonical correlations are ignored at this point and not stored (hence the placeholder ``\_''). A counter $c_k$ for each canonical component is initialised as 0.

The core part of the algorithm are the two loops that run over the permutations in $\mathcal{P}$ and the $K$ canonical components (between lines 25 and 35). At each permutation $j$, \textsc{cca} is executed $K$ times. In each, the columns of $\mathbf{U}$ and $\mathbf{V}$ that precede the current $k$ are removed, such that their respective variances are not allowed to influence the canonical correlations at position $k$. At each permutation, the $K$ canonical correlations are obtained (the third output from the function ``cca'') and used to compute the associated test statistic. As shown, Wilks' statistic, $\lambda_{k}$, is used, simplified by the removal of the constant term, which does not affect permutation p-values. For numerical stability, sum of logarithms is favoured over the logarithm of a product (compare line 30 with Equation \ref{eqn:Wilks}). For inference using Roy's statistic, replace the condition $(\lambda_{k})_{j}^{*} \geqslant \lambda_{k}$ for $(r_{k})_{j}^{*} \geqslant r_{k}$ in line 31; this modification alone is sufficient as $\theta_k$ is permutationally equivalent to $r_k$. In that case, computations indicated in line 30 are no longer needed and can be removed to save computational time.

Whenever the statistic for the correlation at position $k$ in a given permutation is higher or equal than that for the unpermuted data, the counter $c_k$ is incremented (line 32). After the loop, the counter is converted into a p-value for each $k$. These simple, uncorrected p-values, however, are not useful. Instead, \textsc{fwer}-adjusted p-values are computed under closure using the cumulative maximum, i.e., the p-value for $r_k$ is the largest (least significant) uncorrected p-value up to position $k$. The algorithm returns then these adjusted p-values, which can be compared to a predefined test level $\alpha$ to establish significance. Note that $\alpha$ itself is never used in the algorithm.

As presented, the algorithm does not cover dimensionality reduction or any penalty to enforce sparse solutions for \textsc{cca}. Dimensionality reduction using methods such as principal component analysis (\textsc{pca}), if included, would be performed after residualisation, but before \textsc{cca}. Thus, in the algorithm, \textsc{pca} or \textsc{ica}, if executed, would be done between lines 18 and 19. As for the many forms of sparse or penalised \textsc{cca} \citep{Nielsen2002, Waaijenborg2007, Wiesel2008, Parkhomenko2007, Parkhomenko2009, Witten2009, Soneson2010, Hardoon2011, Gao2017, Ma2018, Tan2018}, in principle these can be incorporated into the algorithm through the replacement of the classical \textsc{cca} in lines 19 and 29 for one of these methods.

\section{Evaluation Methods}
\label{sec:evaluation}

In this section we describe the synthetic data and methods used to investigate error rates and power under the different choices for the various aspects presented in Section \ref{sec:theory} at each stage of a permutation test for \textsc{cca}, providing empirical evidence for the approach proposed. An overview of these aspects and choices at each stage is shown in Table \ref{tab:choices}. For each case, we use a series of simulation scenarios: each consists of a set of synthetic variables constructed using random values drawn from a normal or a non-normal (kurtotic or binary) probability distribution, sometimes with or without dimensionality reduction using principal components analysis \citep[\textsc{pca};][]{Hotelling1933, Jolliffe2002}, sometimes with or without signal, and sometimes with or without nuisance variables. We also consider cases with large sample sizes and large number of variables. An overview of these scenarios (there are twenty of them) is in Table \ref{tab:scenarios}.

\begin{sidewaystable}
\caption{A valid permutation test for \textsc{cca} needs to consider several computational and statistical aspects, which can be seen as steps in a procedure; each of them may be addressed using different strategies, some of which are studied below. The recommendations for the different options for each case (column ``Use'') are based on theoretical grounds shown in the sections as indicated in the table, and verified empirically using synthetic data (Section \ref{sec:results}).}
\begin{center}
\begin{tabular}{
@{}
m{50mm}<{\raggedright}
m{89mm}<{\raggedright}
c
m{23mm}<{\raggedright}
m{23mm}<{\raggedright}
@{}}
\toprule
Step & 
Possible strategies studied & 
Use & 
Theory &
Scenarios\\
\midrule
\multirow[c]{2}{=}{\parbox{50mm}{\raggedright Estimation of the canonical components}} & 
(\emph{a}) All in a single step. &
{\color{red}\ding{'67}} &
\multirow[c]{2}{=}{\parbox{23mm}{\raggedright Section \ref{sec:theory:permutation}.}} & 
\multirow[c]{2}{=}{\parbox{23mm}{\raggedright \textsc{i}--\textsc{vi}.}}\\
{} & 
(\emph{b}) Stepwise; variance already explained removed. &
{\color{blue}\ding{'63}} &
{} \\
\midrule
\multirow[c]{2}{=}{\parbox{50mm}{\raggedright Inclusion of the complement of the canonical coefficients}} &
(\emph{a}) Null space not included. &
{\color{red}\ding{'67}} &
\multirow[c]{2}{=}{\parbox{23mm}{\raggedright Section \ref{sec:theory:permutation}.}} &
\multirow[c]{2}{=}{\parbox{23mm}{\raggedright \textsc{i}--\textsc{vi}.}}\\
{} & 
(\emph{b}) Null space included. & 
{\color{blue}\ding{'63}} &
{} \\
\midrule
\multirow[c]{3}{=}{\parbox{50mm}{\raggedright Correction for multiple testing}} & 
(\emph{a}) Uncorrected, simple p-values, $[p_k]_{\text{unc}}$. &
{\color{red}\ding{'67}} &
\multirow[c]{3}{=}{\parbox{23mm}{\raggedright Section \ref{sec:theory:multiplicity}.}}&
\multirow[c]{3}{=}{\parbox{23mm}{\textsc{i}--\textsc{vi}, \textsc{xviii}.}}\\
{} &
(\emph{b}) Corrected, cumulative maximum, $[p_k]_{\text{clo}}$. &
{\color{blue}\ding{'63}} &
{} \\
{} &
(\emph{c}) Corrected, distribution of the maximum, $[p_k]_{\text{max}}$. &
{\color{darkgreen}$\bullet$} &
{} \\
\midrule
\multirow[c]{3}{=}{\parbox{50mm}{\raggedright Treatment of nuisance variables}} &  
(\emph{a}) Simple residualisation ($\mathbf{Q}=\mathbf{I}$). &
{\color{red}\ding{'67}} &
\multirow[c]{3}{=}{\parbox{23mm}{\raggedright Sections \ref{sec:theory:nuisance} and \ref{sec:restricted}.}}&
\multirow[c]{3}{=}{\parbox{23mm}{\textsc{vii}--\textsc{xviii}.}}\\
{} & 
(\emph{b}) Huh--Jhun method. &
{\color{blue}\ding{'63}} &
{} \\
{} &
(\emph{c}) Theil method. &
{\color{blue}\ding{'63}} &
{}\\
\midrule
\multirow[c]{2}{=}{\parbox{50mm}{\raggedright Choice of the test statistic}} & 
(\emph{a}) Wilks' $\lambda_{k}$. &
{\color{blue}\ding{'63}} &
\multirow[c]{2}{=}{\parbox{23mm}{\raggedright Sections \ref{sec:theory:parametric} and \ref{sec:theory:statistic}}} &
\multirow[c]{2}{=}{\parbox{23mm}{\textsc{xvii}, \textsc{xviii}.}}\\
{} & 
(\emph{b}) Roy's largest root, $\theta_{k}$. &
{\color{blue}\ding{'63}} &
{} \\
\bottomrule
\vspace*{.5mm}
\end{tabular}
\begin{tabular}{l@{\hspace{10mm}}l@{\hspace{10mm}}l}
{\color{blue}\ding{'63}} Can or should be used.&
{\color{darkgreen}$\bullet$} Can but should not be used.&
{\color{red}\ding{'67}} Cannot or should not be used.\\
\end{tabular}
\end{center}
\label{tab:choices}
\end{sidewaystable}

\begin{sidewaystable}
\caption{Simulation scenarios.}
\begin{center}
\begin{tabular}{@{}
m{15.5mm}<{\raggedright}  
m{12mm}<{\raggedright}  
m{12mm}<{\raggedright}  
m{11mm}<{\raggedright}  
m{11mm}<{\raggedright}  
m{11mm}<{\raggedright}  
m{11mm}<{\raggedright}  
m{15mm}<{\raggedright}  
m{20mm}<{\raggedright}  
m{15mm}<{\raggedright}  
m{15mm}<{\raggedright}  
m{15mm}<{\raggedright}  
@{}}
\toprule
Scenarios &    & $N$  & $P$ & $Q$ & $R$ & $S$ & \#(\textsc{pca}) & Distribution & Signals & \#Perms. & \#Reps. \\
\midrule
\multirow[c]{6}{=}{\parbox{15mm}{\raggedright Without nuisance}}
   & \textsc{i}     & 100   & 16   & 20   & 0    & 0    & ---  & normal    & ---    & 2000  & 2000\\
{} & \textsc{ii}    & 100   & 16   & 20   & 0    & 0    & 10   & normal    & ---    & 2000  & 2000\\
{} & \textsc{iii}   & 100   & 16   & 20   & 0    & 0    & ---  & kurtotic  & ---    & 2000  & 2000\\
{} & \textsc{iv}    & 100   & 16   & 20   & 0    & 0    & 10   & kurtotic  & ---    & 2000  & 2000\\
{} & \textsc{v}     & 100   & 16   & 20   & 0    & 0    & ---  & binary    & ---    & 2000  & 2000\\
{} & \textsc{vi}    & 100   & 16   & 20   & 0    & 0    & 10   & binary    & ---    & 2000  & 2000\\
\midrule
\multirow[c]{6}{=}{\parbox{15mm}{\raggedright Partial \textsc{cca}}}
   & \textsc{vii}   & 100   & 16   & 20   & 15   & $R$  & ---  & normal    & ---    & 2000  & 2000\\
{} & \textsc{viii}  & 100   & 16   & 20   & 15   & $R$  & 10   & normal    & ---    & 2000  & 2000\\
{} & \textsc{ix}    & 100   & 16   & 20   & 15   & $R$  & ---  & kurtotic  & ---    & 2000  & 2000\\
{} & \textsc{x}     & 100   & 16   & 20   & 15   & $R$  & 10   & kurtotic  & ---    & 2000  & 2000\\
{} & \textsc{xi}    & 100   & 16   & 20   & 15   & $R$  & ---  & binary    & ---    & 2000  & 2000\\
{} & \textsc{xii}   & 100   & 16   & 20   & 15   & $R$  & 10   & binary    & ---    & 2000  & 2000\\
\midrule
\multirow[c]{2}{=}{\parbox{15mm}{\raggedright Bipartial \textsc{cca}}}
   & \textsc{xiii}  & 100   & 16   & 20   & 15   & 15   & ---  & normal    & ---    & 2000  & 2000\\
{} & \textsc{xiv}   & 100   & 16   & 20   & 15   & 15   & 10   & normal    & ---    & 2000  & 2000\\
\midrule
\multirow[c]{2}{=}{\parbox{15mm}{\raggedright Larger samples}}
   & \textsc{xv}    & $*$   & 16   & 20   & 20   & $R$  & ---  & normal    & ---    & 1000  & 1000\\
{} & \textsc{xvi}   & $*$   & 16   & 20   & 20   & $R$  & 10   & normal    & ---    & 1000  & 1000\\
\midrule
\multirow[c]{2}{=}{\parbox{15mm}{\raggedright With signal}}
   & \textsc{xvii}   & 100   & 16   & 20   & 0    & 0    & ---  & normal    & sparse & 2000  & 2000\\
{} & \textsc{xviii}    & 100   & 16   & 20   & 0    & 0    & ---  & normal    & dense  & 2000  & 2000\\
\bottomrule
\end{tabular}
\end{center}
{\footnotesize
* For scenarios \textsc{xv} and \textsc{xvi}, sample size varied, $N=\{100, 200, \ldots , 900, 1000\}$. In the table, $R$ and $S$ refer to the number of nuisance variables other than the intercept, which is always included (so the number of nuisance variables in left and right sides for all the simulation scenarios was always, respectively $R+1$ and $S+1$). For partial \textsc{cca}, the number of nuisance variables on one side is always the same as in the other, i.e., $S=R$, but that does not have to be for bipartial \textsc{cca}, even though here the same size was used. The case with larger samples was used for investigation of partial \textsc{cca}. 
\par}
\label{tab:scenarios}
\end{sidewaystable}

We start by investigating aspects related to the estimation of the canonical components at each permutation. Specifically, we consider (\emph{a}) a one-step estimation of all canonical components, from $1$ to $K$, versus (\emph{b}) sequential estimation that removes, for the $k$-th canonical component in a given permutation, the variance already explained by the previous ones, as described in Section \ref{sec:theory:permutation}. With respect to the inclusion of the complement of the canonical coefficients, we consider (\emph{a}) without the inclusion of the null space of the canonical coefficients, versus (\emph{b}) with its inclusion so as to ensure that all variance from the original data not explained in the initial \textsc{cca} is considered in the estimation at every permutation, as described in Section \ref{sec:theory:permutation}. With respect to multiple testing, we consider the following strategies: (\emph{a}) simple, uncorrected p-values, $[p_k]_{\text{unc}}$, (\emph{b}) corrected under closure, $[p_k]_{\text{clo}}$, and (\emph{c}) corrected using the distribution of the maximum statistic $[p_k]_{\text{max}}$; both $[p_k]_{\text{clo}}$ and $[p_k]_{\text{max}}$ offer \textsc{fwer} control, as discussed in Section \ref{sec:theory:multiplicity}. Keeping the same notation, we define scenarios \textsc{i}--\textsc{vi} consisting of $N=100$ observations, with $P=16$ variables on the left side ($\mathbf{Y}$) of \textsc{cca} and $Q=20$ variables on the right side ($\mathbf{X}$) (the procedure is symmetric; the choice of sides is arbitrary and does not affect results); for these six scenarios, data are drawn from one of three possible distributions: a normal distribution with zero mean and unit variance, a Student's $t$ distribution with variable degrees of freedom $\nu = \left\{2, 4, 6, 8, 10\right\}$ (kurtotic), or a Bernoulli distribution with parameter $q=0.20$ (binary). Analyses with and without dimensionality reduction to 10 variables using \textsc{pca} are considered. The number of permutations used to compute p-values was set as $J=2000$, with $2000$ realisations (repetitions), thus allowing the computation of error rates.

We then turn our attention to aspects related to nuisance and residualisation discussed in Sections \ref{sec:theory:nuisance} and \ref{sec:restricted}. We consider (\emph{a}) simple residualisation, (\emph{b}) residualisation using the Huh--Jhun method, and (\emph{c}) residualisation using the Theil method. For this purpose, scenarios \textsc{vii}--\textsc{xvii} are constructed similarly as \textsc{i}--\textsc{vi}, except that a third set $\mathbf{Z}$ of $R=15$ variables is used as nuisance for partial \textsc{cca}, whereas two other scenarios, \textsc{xiii} and \textsc{xiv}, a fourth set of variables $\mathbf{W}$ of $S=15$ variables is used as nuisance for bipartial \textsc{cca}.

The impact of ignoring, in samples substantially larger than the number of variables, the dependencies introduced by the residualisation of both sides of \textsc{cca} is studied with scenarios \textsc{xv} and \textsc{xvi}, that consider samples progressively larger, $N=\left\{100, 200, \ldots, 900, 1000\right\}$, while keeping the other parameters similar as in scenarios \textsc{vii} and \textsc{viii}. Finally, we briefly investigate power and the choice of the test statistic: we consider (\emph{a}) Wilks' statistic ($\lambda_{k}$), as well as (\emph{b}) Roy's largest root ($\theta_{k}$), as discussed in Sections \ref{sec:theory:parametric} and \ref{sec:theory:statistic}. We define scenarios \textsc{xvii} and \textsc{xviii} similarly as \textsc{i}, this time including a strong, true signal in one canonical component, thus named ``sparse'', or multiple, weaker signals shared across multiple (half of the smaller set, thus, ``dense''). For all scenarios, an intercept is always included as nuisance variable in both sides such that the actual number of nuisance variables is $R+1$ and $S+1$ for each side, respectively. To report confidence intervals (95\%), the \citet{Wilson1927} method is used.

\section{Results}
\label{sec:results}

In the results below, the Sections \ref{sec:results:estimation} and \ref{sec:results:multiplicity} establish empirically that with an estimation method (i) that includes the null space of the canonical coefficients, (ii) that finds the canonical correlations in an iterative manner, and (iii) that after computing p-values through a closed testing procedure, the error rates are controlled. The subsequent results, from Section \ref{sec:results:nuisance} onwards, consider only this valid approach.

\subsection{Estimation strategies}
\label{sec:results:estimation}

Not including the complement of the canonical coefficients (null space) caused error rates to be dramatically inflated, well above the expected test level $\alpha=0.05$ (5\%), regardless of whether the estimation used the single step or the stepwise procedure, and regardless of any of the multiple testing correction strategies discussed; these results are shown in Table \ref{tab:results:estimation+mtp}.

\begin{table}[!p]
\caption{Observed per comparison error rate (\%) and 95\% confidence intervals for the first 6 canonical correlations in scenario \textsc{i}, assessed using the Wilks' statistic and three different multiple testing correction methods; the observed familywise error rate (\textsc{fwer}) for each case is also shown. Valid methods should have a \textsc{fwer} close to the nominal 5\%, and \textsc{pcer} close to the nominal $(5\%)^k$; see Sections \ref{sec:results:estimation} and \ref{sec:results:multiplicity} for details.}
\begin{center}
\hspace*{-.75cm}
\begin{tabular}{@{}m{12mm}<{\raggedright}m{29mm}<{\raggedright}m{29mm}<{\raggedright}m{29mm}<{\raggedright}m{24mm}<{\raggedright}@{}}
\toprule
{} & \multicolumn{2}{l}{Null space not included} & \multicolumn{2}{l}{Null space included}\\
\cmidrule(lr){2-3} \cmidrule(l){4-5}
{} & Single step & Stepwise & Single step & Stepwise \\
\midrule
\multicolumn{5}{l}{(\emph{a}) \textit{Uncorrected, simple p-values,} $[p_k]_{\text{unc}}$\textit{.}}\\
$k=1$ & 
91.35 \scalebox{.7}[1]{(90.04--92.50)} & 
91.35 \scalebox{.7}[1]{(90.04--92.50)} & 
4.70 \scalebox{.7}[1]{(3.86--5.72)} & 
4.70 \scalebox{.7}[1]{(3.86--5.72)} \\
$k=2$ & 
93.50 \scalebox{.7}[1]{(92.33--94.50)} & 
60.40 \scalebox{.7}[1]{(58.24--62.52)} & 
4.60 \scalebox{.7}[1]{(3.77--5.61)} & 
0.25 \scalebox{.7}[1]{(0.11--0.58)} \\
$k=3$ & 
94.70 \scalebox{.7}[1]{(93.63--95.60)} & 
26.70 \scalebox{.7}[1]{(24.81--28.68)} & 
4.60 \scalebox{.7}[1]{(3.77--5.61)} & 
0.00 \scalebox{.7}[1]{(0.00--0.19)} \\
$k=4$ & 
95.55 \scalebox{.7}[1]{(94.56--96.37)} & 
7.25 \scalebox{.7}[1]{(6.19--8.47)} & 
4.85 \scalebox{.7}[1]{(3.99--5.88)} & 
0.00 \scalebox{.7}[1]{(0.00--0.19)} \\
$k=5$ & 
96.10 \scalebox{.7}[1]{(95.16--96.86)} & 
1.45 \scalebox{.7}[1]{(1.01--2.07)} & 
4.40 \scalebox{.7}[1]{(3.59--5.39)} & 
0.00 \scalebox{.7}[1]{(0.00--0.19)} \\
$k=6$ & 
96.75 \scalebox{.7}[1]{(95.88--97.44)} & 
0.25 \scalebox{.7}[1]{(0.11--0.58)} & 
4.30 \scalebox{.7}[1]{(3.50--5.28)} & 
0.00 \scalebox{.7}[1]{(0.00--0.19)} \\
\textsc{fwer} &
99.90 \scalebox{.7}[1]{(99.64--99.97)} & 
91.35 \scalebox{.7}[1]{(90.04--92.50)} & 
18.30 \scalebox{.7}[1]{(16.67--20.05)} & 
4.70 \scalebox{.7}[1]{(3.86--5.72)} \\
\midrule
\multicolumn{5}{l}{(\emph{b}) \textit{Corrected, cumulative maximum,} $[p_k]_{\text{clo}}$\textit{.}}\\
$k=1$ & 
91.35 \scalebox{.7}[1]{(90.04--92.50)} & 
91.35 \scalebox{.7}[1]{(90.04--92.50)} & 
4.70 \scalebox{.7}[1]{(3.86--5.72)} & 
4.70 \scalebox{.7}[1]{(3.86--5.72)} \\
$k=2$ & 
90.35 \scalebox{.7}[1]{(88.98--91.57)} & 
60.40 \scalebox{.7}[1]{(58.24--62.52)} & 
3.40 \scalebox{.7}[1]{(2.69--4.29)} & 
0.25 \scalebox{.7}[1]{(0.11--0.58)} \\
$k=3$ & 
89.80 \scalebox{.7}[1]{(88.40--91.05)} & 
26.70 \scalebox{.7}[1]{(24.81--28.68)} & 
2.75 \scalebox{.7}[1]{(2.12--3.56)} & 
0.00 \scalebox{.7}[1]{(0.00--0.19)} \\
$k=4$ & 
89.55 \scalebox{.7}[1]{(88.13--90.82)} & 
7.25 \scalebox{.7}[1]{(6.19--8.47)} & 
2.40 \scalebox{.7}[1]{(1.81--3.17)} & 
0.00 \scalebox{.7}[1]{(0.00--0.19)} \\
$k=5$ & 
89.30 \scalebox{.7}[1]{(87.87--90.58)} & 
1.45 \scalebox{.7}[1]{(1.01--2.07)} & 
1.75 \scalebox{.7}[1]{(1.26--2.42)} & 
0.00 \scalebox{.7}[1]{(0.00--0.19)} \\
$k=6$ & 
88.85 \scalebox{.7}[1]{(87.40--90.16)} & 
0.25 \scalebox{.7}[1]{(0.11--0.58)} & 
1.45 \scalebox{.7}[1]{(1.01--2.07)} & 
0.00 \scalebox{.7}[1]{(0.00--0.19)} \\
\textsc{fwer} &
91.35 \scalebox{.7}[1]{(90.04--92.50)} & 
91.35 \scalebox{.7}[1]{(90.04--92.50)} & 
4.70 \scalebox{.7}[1]{(3.86--5.72)} & 
4.70 \scalebox{.7}[1]{(3.86--5.72)} \\
\midrule
\multicolumn{5}{l}{(\emph{c}) \textit{Corrected, distribution of the maximum,} $[p_k]_{\text{max}}$\textit{.}}\\
$k=1$ & 
91.35 \scalebox{.7}[1]{(90.04--92.50)} & 
91.35 \scalebox{.7}[1]{(90.04--92.50)} & 
4.70 \scalebox{.7}[1]{(3.86--5.72)} & 
4.70 \scalebox{.7}[1]{(3.86--5.72)} \\
$k=2$ & 
7.95 \scalebox{.7}[1]{(6.84--9.22)} & 
10.95 \scalebox{.7}[1]{(9.66--12.39)} & 
0.00 \scalebox{.7}[1]{(0.00--0.19)} & 
0.00 \scalebox{.7}[1]{(0.00--0.19)} \\
$k=3$ & 
0.00 \scalebox{.7}[1]{(0.00--0.19)} & 
0.00 \scalebox{.7}[1]{(0.00--0.19)} & 
0.00 \scalebox{.7}[1]{(0.00--0.19)} & 
0.00 \scalebox{.7}[1]{(0.00--0.19)} \\
$k=4$ & 
0.00 \scalebox{.7}[1]{(0.00--0.19)} & 
0.00 \scalebox{.7}[1]{(0.00--0.19)} & 
0.00 \scalebox{.7}[1]{(0.00--0.19)} & 
0.00 \scalebox{.7}[1]{(0.00--0.19)} \\
$k=5$ & 
0.00 \scalebox{.7}[1]{(0.00--0.19)} & 
0.00 \scalebox{.7}[1]{(0.00--0.19)} & 
0.00 \scalebox{.7}[1]{(0.00--0.19)} & 
0.00 \scalebox{.7}[1]{(0.00--0.19)} \\
$k=6$ & 
0.00 \scalebox{.7}[1]{(0.00--0.19)} & 
0.00 \scalebox{.7}[1]{(0.00--0.19)} & 
0.00 \scalebox{.7}[1]{(0.00--0.19)} & 
0.00 \scalebox{.7}[1]{(0.00--0.19)} \\
\textsc{fwer} &
91.35 \scalebox{.7}[1]{(90.04--92.50)} & 
91.35 \scalebox{.7}[1]{(90.04--92.50)} & 
4.70 \scalebox{.7}[1]{(3.86--5.72)} & 
4.70 \scalebox{.7}[1]{(3.86--5.72)} \\
\bottomrule
\end{tabular}
\end{center}
{\footnotesize
Using the Roy's statistic led to similar results as with Wilks (not shown). Dimensionality reduction with \textsc{pca} led to similar results for the case in which the null space is included (not shown). For the case in which the null space is not included, results are not comparable with the ones above because, after \textsc{pca}, $P=Q$ in the simulations, such that there is no null space to be considered as the matrices with canonical coefficients in both sides are then square.\par}
\label{tab:results:estimation+mtp}
\end{table}

Even when the null space of the canonical coefficients was included, a single step procedure was never satisfactory. To understand this, consider the following consequence of the theory presented in Sections \ref{sec:theory:parametric} and \ref{sec:theory:multiplicity}: for a valid, exact test in \textsc{cca}, the expected error rate for each $\mathcal{H}^0_k$, i.e., the \textit{per comparison error rate} \citep[\textsc{pcer};][]{Hochberg1987} is $\alpha$ for $k=1$, but for $k = 2$ it is $\alpha \times \alpha$, since the null can only be rejected if the previous one has also been declared significant at $\alpha$. More generally, the \textsc{pcer} for a valid test is $\alpha^k$ for the $k$-th test, i.e., for the $k$-th canonical correlation. If the test level is set at 5\%, then the \textsc{pcer} is 5\% for $k=1$, 0.25\% for $k=2$, 0.0125\% for $k=3$, and so forth. Error rates above this expectation render the test invalid; below render it conservative. In the simulations, a single step procedure never led to an exact test, with or without consideration to multiple testing, as shown in Table \ref{tab:results:estimation+mtp}.

\subsection{Multiple testing}
\label{sec:results:multiplicity}

As with the \textsc{pcer}, it is worth mentioning what the expected \textsc{fwer} for a valid, exact test is. That expectation is the test level itself, i.e., $\alpha$. Any higher error rate renders a test invalid; lower error rate renders it conservative, though valid. Table \ref{tab:results:estimation+mtp} shows the \textsc{fwer} for the three different correction methods considered.

\begin{figure}[!p]
\centering
\includegraphics[width=\textwidth]{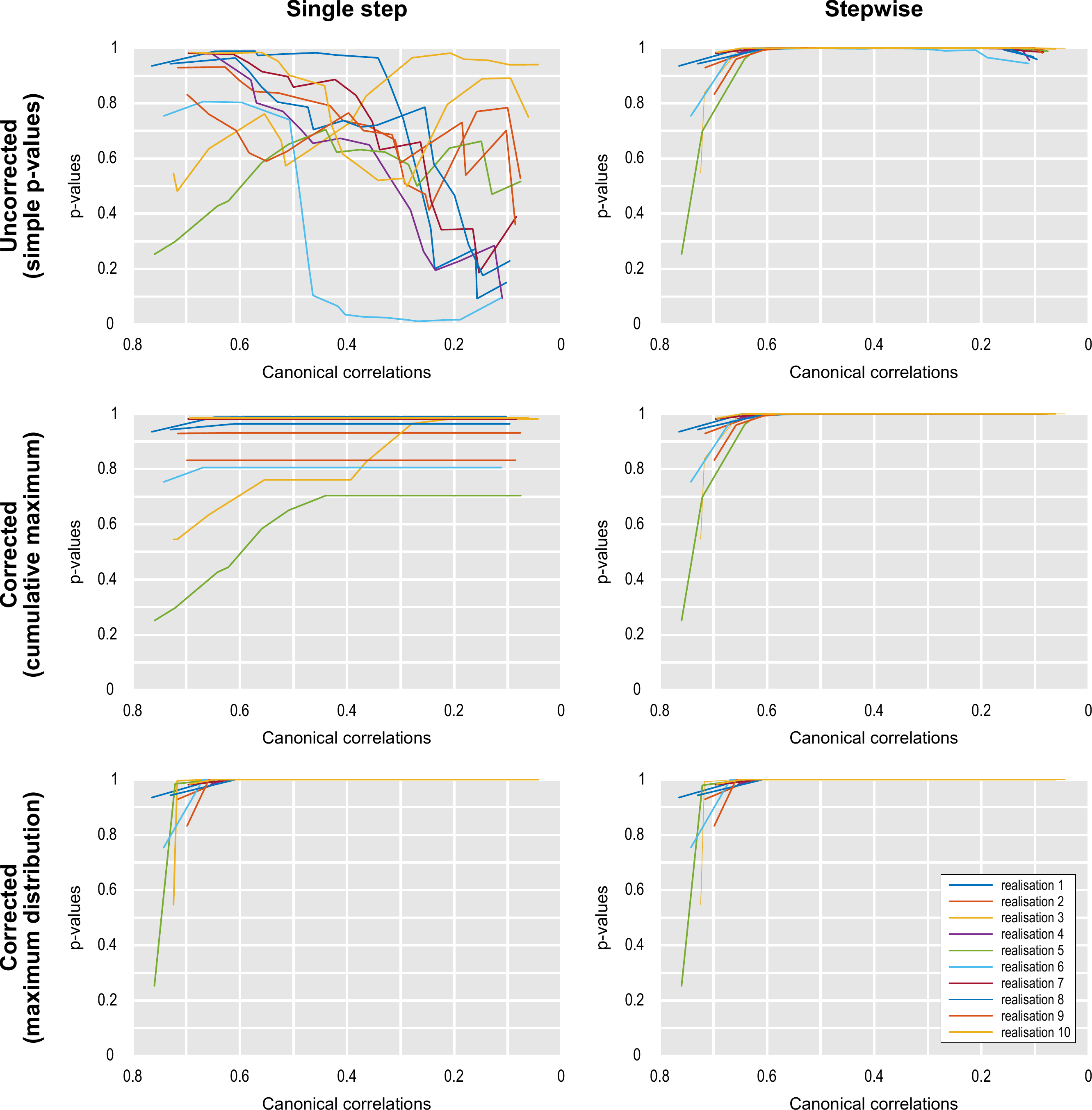}
\caption{Relationship between canonical correlations (horizontal axes) and associated p-values (vertical axes) for 10 realisations of scenario \textsc{i}, considering two estimation methods (single step and stepwise) and three multiple testing correction methods (uncorrected, corrected using the cumulative maximum, and corrected using the distribution of the maximum statistic). The figure complements Table \ref{tab:results:estimation+mtp} by showing example realisations that average to the error rates shown in the table for the cases in which the null space is included. For simple, uncorrected p-values, the test is inadmissible; for corrected using the distribution of the maximum statistic, the test is overly conservative; single step does not control the familywise error rate.}
\label{fig:pvals}
\end{figure}

If the null space was not included, since the \textsc{pcer} was not controlled, the \textsc{fwer} could not be controlled either (first two columns). If the null space of the canonical coefficients was included (last two columns), even though the single step estimation controlled the \textsc{pcer}, the \textsc{fwer} was not controlled for the simple, uncorrected p-values (third column, upper panel), which is not surprising. It should be emphasised, however, that these simple p-values have another problem: they are not guaranteed to be monotonically related to the respective canonical correlations, such that it is possible that, using these p-values, the null hypothesis could be rejected for some canonical correlation, but retained for another that happens to be larger than the former. The use of such uncorrected, simple p-values, therefore, constitutes a test that is inadmissible. The problem with lack of monotonicity with uncorrected p-values is less severe if estimation is done in a stepwise manner (fourth column, upper panel), but is nonetheless still present, as shown in Figure \ref{fig:pvals}, and has potential to lead to an excess \textsc{fwer}, even though that did not occur in these simulations.

For the other two correction methods, when the null space of the canonical coefficients was included in the estimation process, \textsc{fwer} was controlled (third and fourth columns of Table \ref{tab:results:estimation+mtp}, middle and lower panels), but there are particularities. Using the distribution of the maximum (lower panel) led to very conservative \textsc{pcer}, for both single step or stepwise estimation, whereas correction with closure led to invalid \textsc{pcer} for single step estimation (third column, middle panel).

The only configuration that led to exact (neither conservative or invalid) control over \textsc{pcer} and \textsc{fwer}, and a monotonic relationship between canonical correlations and associated p-values, is the one in which a stepwise estimation was performed, with the null space of the canonical coefficients included, and with correction using a closed testing procedure (fourth column, middle panel of Table \ref{tab:results:estimation+mtp}). Moreover, the \textsc{fwer}, when controlled using the cumulative maximum or the distribution of the maximum statistic, is guaranteed to match the \textsc{pcer} for $k=1$: in the former case, any further rejection of the null is conditional on the first one having been rejected; in the latter, the distribution of the maximum coincides with the distribution of the first as the canonical correlations are ranked from largest to smallest.

\subsection{Nuisance variables}
\label{sec:results:nuisance}

For partial \textsc{cca}, simple residualisation, even using the above procedure (stepwise estimation, null space included, correction via closure), resulted in the error rates being dramatically inflated, as shown in Table \ref{tab:results:nuisance}. The Huh--Jhun and the Theil residualisation methods, in contrast, resulted in the error rates being controlled at the nominal level, with no excess false positives. For bipartial \textsc{cca}, the problem did not happen in the simulation settings: simple residualisation of both sides by entirely different sets of variables did not cause the error rates to be inflated; yet, using Huh-Jhun or Theil also produced nominal error rates, suggesting that these could be used in any configuration of nuisance variables, regardless of whether those in one side are not independent from those in the other.

\begin{table}[t]
\caption{Observed per comparison error rate (\textsc{pcer}, \%) and 95\% confidence intervals for the first 6 canonical correlations in scenarios \textsc{vii} (partial \textsc{cca}) and \textsc{xiii} (bipartial \textsc{cca}), for the three different methods considered for treatment of nuisance variables.}
\begin{center}
\begin{tabular}{@{}m{25mm}<{\raggedright}m{32mm}<{\raggedright}m{32mm}<{\raggedright}m{25mm}<{\raggedright}@{}}
\toprule
{} & Simple residuals & Huh--Jhun & Theil\\
\midrule
\multicolumn{4}{l}{(\emph{a}) \textit{Partial \textsc{cca}}}\\
$k=1$ (\textsc{fwer}) & 
83.85 \scalebox{.7}[1]{(82.17--85.40)} & 
5.10 \scalebox{.7}[1]{(4.22--6.15)} & 
4.85 \scalebox{.7}[1]{(3.99--5.88)} \\
$k=2$ & 
44.15 \scalebox{.7}[1]{(41.99--46.34)} & 
0.30 \scalebox{.7}[1]{(0.14--0.65)} & 
0.35 \scalebox{.7}[1]{(0.17--0.72)} \\
$k=3$ & 
12.75 \scalebox{.7}[1]{(11.36--14.28)} & 
0.05 \scalebox{.7}[1]{(0.01--0.28)} & 
0.00 \scalebox{.7}[1]{(0.00--0.19)} \\
$k=4$ & 
1.75 \scalebox{.7}[1]{(1.26--2.42)} & 
0.00 \scalebox{.7}[1]{(0.00--0.19)} & 
0.00 \scalebox{.7}[1]{(0.00--0.19)} \\
$k=5$ & 
0.20 \scalebox{.7}[1]{(0.08--0.51)} & 
0.00 \scalebox{.7}[1]{(0.00--0.19)} & 
0.00 \scalebox{.7}[1]{(0.00--0.19)} \\
$k=6$ & 
0.00 \scalebox{.7}[1]{(0.00--0.19)} & 
0.00 \scalebox{.7}[1]{(0.00--0.19)} & 
0.00 \scalebox{.7}[1]{(0.00--0.19)} \\
\midrule
\multicolumn{4}{l}{(\emph{b}) \textit{Bipartial \textsc{cca}}}\\
$k=1$ (\textsc{fwer}) & 
5.55 \scalebox{.7}[1]{(4.63--6.64)} & 
5.20 \scalebox{.7}[1]{(4.31--6.26)} & 
4.45 \scalebox{.7}[1]{(3.63--5.44)} \\
$k=2$ & 
0.10 \scalebox{.7}[1]{(0.03--0.36)} & 
0.30 \scalebox{.7}[1]{(0.14--0.65)} & 
0.20 \scalebox{.7}[1]{(0.08--0.51)} \\
$k=3$ & 
0.00 \scalebox{.7}[1]{(0.00--0.19)} & 
0.00 \scalebox{.7}[1]{(0.00--0.19)} & 
0.00 \scalebox{.7}[1]{(0.00--0.19)} \\
$k=4$ & 
0.00 \scalebox{.7}[1]{(0.00--0.19)} & 
0.00 \scalebox{.7}[1]{(0.00--0.19)} & 
0.00 \scalebox{.7}[1]{(0.00--0.19)} \\
$k=5$ & 
0.00 \scalebox{.7}[1]{(0.00--0.19)} & 
0.00 \scalebox{.7}[1]{(0.00--0.19)} & 
0.00 \scalebox{.7}[1]{(0.00--0.19)} \\
$k=6$ & 
0.00 \scalebox{.7}[1]{(0.00--0.19)} & 
0.00 \scalebox{.7}[1]{(0.00--0.19)} & 
0.00 \scalebox{.7}[1]{(0.00--0.19)} \\
\bottomrule
\end{tabular}
\end{center}
{\footnotesize
Estimation included the null space of the canonical coefficients and a stepwise procedure, assessed using the Wilks' statistic, and corrected using a closed testing procedure (\textsc{ctp}). The \textsc{ctp} guarantees that the familywise error rate (\textsc{fwer}) matches the \textsc{pcer} for the first canonical correlation (i.e., for $k=1$). Using the Roy's statistic led to similar results as with Wilks'; likewise, dimensionality reduction with \textsc{pca} led to similar results (not shown).\par}
\label{tab:results:nuisance}
\end{table}

\subsection{Non-normality}
\label{sec:results:normality}

Without nuisance variables and with kurtotic data simulated using a Student's $t$ distribution with a small number of degrees of freedom, $\nu = \left\{2, 4, 6, 8, 10\right\}$, as well as with binary data simulated using a Bernoulli distribution with parameter $q=0.20$, error rates were controlled nominally, as shown in Table \ref{tab:results:normality+partial}. In partial \textsc{cca}, however, even using the Huh--Jhun method, highly kurtotic data led to excess error rates. In particular, for the simulated data using a Student's $t$ distribution with degrees of freedom of only $\nu=2$, the observed error rate was 14.7\%, for a test level of 5\%; using the Theil method led to also inflated error rate in this case, with 10.7\% (95\% confidence interval: 8.28--13.72, not shown in the table). For $\nu \geqslant 4$, error rates were controlled at the nominal level, for both Huh--Jhun (Table \ref{tab:results:normality+partial}) and Theil (not shown).

\begin{table}[!t]
\caption{Observed per comparison error rate (\textsc{pcer}, \%) and 95\% confidence intervals for the first canonical correlation in scenarios \textsc{i}--\textsc{vi} (without nuisance) and \textsc{vi}--\textsc{xii} (partial \textsc{cca}, with Huh--Jhun), considering different distributions for the data.}
\begin{center}
\begin{tabular}{@{}m{25mm}<{\raggedright}m{29mm}<{\raggedright}m{30mm}<{\raggedright}m{30mm}<{\raggedright}@{}}
\toprule
\multicolumn{2}{l}{\!\!\!Distribution} & Without nuisance & Partial \textsc{cca}\\
\midrule
Normal & 
{} & 
4.70 \scalebox{.7}[1]{(3.86--5.72)} & 
5.15 \scalebox{.7}[1]{(4.26--6.21)} \\
\multirow[c]{5}{=}{\parbox{35mm}{\raggedright Student\hspace{10mm}$\begin{cases}{}\\[30pt]{}\end{cases}$}} &
$\nu=2$ & 
3.95 \scalebox{.7}[1]{(3.18--4.90)} & 
14.70 \scalebox{.7}[1]{(13.22--16.32)} \\
{} & 
$\nu=4$ & 
5.45 \scalebox{.7}[1]{(4.54--6.53)} & 
5.40 \scalebox{.7}[1]{(4.49--6.48)} \\
{} & 
$\nu=6$ & 
4.15 \scalebox{.7}[1]{(3.36--5.12)} & 
5.40 \scalebox{.7}[1]{(4.49--6.48)} \\
{} & 
$\nu=8$ & 
4.70 \scalebox{.7}[1]{(3.86--5.72)} & 
5.00 \scalebox{.7}[1]{(4.13--6.04)} \\
{} & 
$\nu=10$ & 
3.85 \scalebox{.7}[1]{(3.09--4.79)} & 
5.10 \scalebox{.7}[1]{(4.22--6.15)} \\
Bernoulli & 
$q=0.2$ & 
5.30 \scalebox{.7}[1]{(4.40--6.37)} & 
5.30 \scalebox{.7}[1]{(4.40--6.37)} \\
\bottomrule
\end{tabular}
\end{center}
{\footnotesize
$\nu$: Degrees of freedom of the Student's $t$ distribution used to simulate data; 
$q$: Parameter of the Bernoulli distribution used to simulate data.
Estimation used the null space of the canonical coefficients and a stepwise procedure, assessed using the Wilks' statistic, and corrected using a closed testing procedure (\textsc{ctp}). The \textsc{ctp} guarantees that the familywise error rate (\textsc{fwer}) matches the \textsc{pcer} for the first canonical correlation (i.e., for $k=1$). Using the Roy's statistic led to similar results as with Wilks'; using Theil led to similar results as Huh--Jhun; likewise, dimensionality reduction with \textsc{pca} led to similar results (not shown).\par}
\label{tab:results:normality+partial}
\end{table}

\subsection{Large samples}
\label{sec:results:large}

Increasing the sample size while keeping the number of variables fixed progressively reduced the amount of errors for the simple residualisation method to treat nuisance variables, as shown in Table \ref{tab:results:large}; the trend was similar with or without dimensionality reduction using \textsc{pca}. The reduction in the error rate as the sample size increased did not affect Huh--Jhun or Theil methods, for which error rates were already controlled even with a relatively smaller sample compared to the number of variables.

\begin{sidewaystable}
\caption{Observed familywise error rate (\textsc{fwer}, \%, that matches the \textsc{pcer} for $k=1$) and 95\% confidence intervals for scenarios \textsc{xv} and \textsc{xvi}, used to investigate effect of different sample sizes, for the three different methods for dealing with nuisance variables.}
\begin{center}
\begin{tabular}{@{}m{10mm}<{\raggedright}m{30mm}<{\raggedright}m{30mm}<{\raggedright}m{30mm}<{\raggedright}m{30mm}<{\raggedright}m{30mm}<{\raggedright}m{24mm}<{\raggedright}@{}}
\toprule
{} & \multicolumn{3}{l}{Without \textsc{pca}} & \multicolumn{3}{l}{With \textsc{pca}}\\
\cmidrule(lr){2-4} \cmidrule(l){5-7}
$N$ & Simple residuals & Huh--Jhun & Theil & Simple residuals & Huh--Jhun & Theil\\
\midrule
100 & 
96.60 \scalebox{.7}[1]{(95.29--97.56)} & 
4.80 \scalebox{.7}[1]{(3.64--6.31)} & 
5.00 \scalebox{.7}[1]{(3.81--6.53)} & 
59.20 \scalebox{.7}[1]{(56.12--62.21)} & 
5.00 \scalebox{.7}[1]{(3.81--6.53)} & 
5.10 \scalebox{.7}[1]{(3.90--6.64)} \\
200 & 
42.20 \scalebox{.7}[1]{(39.17--45.29)} & 
5.50 \scalebox{.7}[1]{(4.25--7.09)} & 
5.40 \scalebox{.7}[1]{(4.16--6.98)} & 
22.30 \scalebox{.7}[1]{(19.83--24.98)} & 
5.00 \scalebox{.7}[1]{(3.81--6.53)} & 
4.50 \scalebox{.7}[1]{(3.38--5.97)} \\
300 & 
25.10 \scalebox{.7}[1]{(22.51--27.88)} & 
5.00 \scalebox{.7}[1]{(3.81--6.53)} & 
5.40 \scalebox{.7}[1]{(4.16--6.98)} & 
12.20 \scalebox{.7}[1]{(10.31--14.37)} & 
5.40 \scalebox{.7}[1]{(4.16--6.98)} & 
5.10 \scalebox{.7}[1]{(3.90--6.64)} \\
400 & 
18.00 \scalebox{.7}[1]{(15.74--20.50)} & 
4.30 \scalebox{.7}[1]{(3.21--5.74)} & 
5.10 \scalebox{.7}[1]{(3.90--6.64)} & 
11.80 \scalebox{.7}[1]{(9.95--13.95)} & 
4.50 \scalebox{.7}[1]{(3.38--5.97)} & 
5.70 \scalebox{.7}[1]{(4.43--7.31)} \\
500 & 
11.50 \scalebox{.7}[1]{(9.67--13.63)} & 
6.10 \scalebox{.7}[1]{(4.78--7.76)} & 
4.10 \scalebox{.7}[1]{(3.04--5.51)} & 
9.50 \scalebox{.7}[1]{(7.83--11.48)} & 
6.80 \scalebox{.7}[1]{(5.40--8.53)} & 
5.10 \scalebox{.7}[1]{(3.90--6.64)} \\
600 & 
10.80 \scalebox{.7}[1]{(9.02--12.88)} & 
5.20 \scalebox{.7}[1]{(3.99--6.76)} & 
5.00 \scalebox{.7}[1]{(3.81--6.53)} & 
9.30 \scalebox{.7}[1]{(7.65--11.26)} & 
4.70 \scalebox{.7}[1]{(3.55--6.19)} & 
4.70 \scalebox{.7}[1]{(3.55--6.19)} \\
700 & 
11.20 \scalebox{.7}[1]{(9.39--13.31)} & 
4.20 \scalebox{.7}[1]{(3.12--5.63)} & 
4.40 \scalebox{.7}[1]{(3.29--5.86)} & 
7.20 \scalebox{.7}[1]{(5.76--8.97)} & 
5.70 \scalebox{.7}[1]{(4.43--7.31)} & 
5.50 \scalebox{.7}[1]{(4.25--7.09)} \\
800 & 
10.10 \scalebox{.7}[1]{(8.38--12.12)} & 
5.50 \scalebox{.7}[1]{(4.25--7.09)} & 
4.20 \scalebox{.7}[1]{(3.12--5.63)} & 
7.00 \scalebox{.7}[1]{(5.58--8.75)} & 
4.80 \scalebox{.7}[1]{(3.64--6.31)} & 
5.90 \scalebox{.7}[1]{(4.60--7.54)} \\
900 & 
8.40 \scalebox{.7}[1]{(6.84--10.28)} & 
5.00 \scalebox{.7}[1]{(3.81--6.53)} & 
4.00 \scalebox{.7}[1]{(2.95--5.40)} & 
7.70 \scalebox{.7}[1]{(6.20--9.52)} & 
5.20 \scalebox{.7}[1]{(3.99--6.76)} & 
5.20 \scalebox{.7}[1]{(3.99--6.76)} \\
1000 & 
7.70 \scalebox{.7}[1]{(6.20--9.52)} & 
4.30 \scalebox{.7}[1]{(3.21--5.74)} & 
5.90 \scalebox{.7}[1]{(4.60--7.54)} & 
5.00 \scalebox{.7}[1]{(3.81--6.53)} & 
6.20 \scalebox{.7}[1]{(4.87--7.87)} & 
4.70 \scalebox{.7}[1]{(3.55--6.19)} \\
\bottomrule
\end{tabular}
\end{center}
{\footnotesize
Without \textsc{pca}: $P=20$, $Q=16$. With \textsc{pca}: $P=Q=10$. Estimation included the null space of the canonical coefficients and a stepwise procedure, assessed using the Wilks' statistic, and corrected using a closed testing procedure (\textsc{ctp}). The \textsc{ctp} guarantees that the familywise error rate (\textsc{fwer}) matches the \textsc{pcer} for the first canonical correlation (i.e., for $k=1$). Using the Roy's statistic led to similar results as with Wilks'. The confidence intervals are wider than for other tables because the number of realisations (and also of permutations) was smaller (Table \ref{tab:scenarios})\par}
\label{tab:results:large}
\end{sidewaystable}

\subsection{Dimensionality reduction}
\label{sec:results:pca}

Dimensionality reduction with \textsc{pca} did not affect error rates (\textsc{pcer} and \textsc{fwer}) with respect to single step vs.\ stepwise estimation of canonical coefficients, nor correction for multiple testing, nor method for addressing nuisance variables. That is, these results (not shown) were indistinguishable from those obtained without \textsc{pca} (shown above). Moreover, as the simulations used the same number of principal components for both sides of \textsc{cca}, including or not the null space could not have affected results, as $P=Q$ after dimensionality reduction. Using \textsc{pca} did yield higher power to detect effects, for both Wilks' and Roy's test statistics (Table \ref{tab:results:power+pca}, next item). This apparent extra power can be attributed to the smaller number of variables after \textsc{pca}, as the principal components that were retained contained most of the simulated signal, which, given the reduced dimensionality of the set of data, could then be detected with higher likelihood.

\subsection{Choice of the statistic}
\label{sec:results:statistic}

The results above, that consider solely the error rates, and are based on results with the Wilks' statistic ($\lambda_{k}$), are essentially the same for Roy's largest root ($\theta_{k}$; results not shown). That is, results regarding the estimation strategies, multiple testing, nuisance variables, non-normality, behaviour with large samples, and dimensionality reduction with \textsc{pca}, are virtually the same for Wilks' and Roy's statistics. In the presence of synthetic signal, however, the two test statistics diverged. Table \ref{tab:results:power+pca} shows that, with signal spread across multiple canonical components (i.e., ``dense''), Wilks' is substantially more powerful than Roy's statistic. With signal concentrated in just one (the first) canonical variable (i.e., ``sparse''), the trend reverses, and Roy's become more powerful than Wilks'.

\begin{sidewaystable}
\caption{Observed power (\%) and 95\% confidence intervals for the first canonical correlation in scenarios \textsc{xvii} and \textsc{xviii}, that included a synthetic signal added to either one or half (``sparse'' or ``dense'', respectively) of the initial variables, thus captured by only one (sparse case) or multiple (dense case) canonical correlations.}
\begin{center}
\begin{tabular}{@{}m{15mm}<{\raggedright}m{15mm}<{\raggedright}m{30mm}<{\raggedright}m{30mm}<{\raggedright}m{30mm}<{\raggedright}m{30mm}<{\raggedright}@{}}
\toprule
{} & {} & \multicolumn{2}{l}{Without \textsc{pca}} & \multicolumn{2}{l}{With \textsc{pca}}\\
\cmidrule(lr){3-4} \cmidrule(l){5-6}
\multicolumn{2}{l}{\!\!\!Signals} & Wilks ($\lambda$) & Roy ($\theta$) & Wilks ($\lambda$) & Roy ($\theta$)\\
\midrule
Sparse & 
$k=1$ & 
42.10 \scalebox{.7}[1]{(39.95--44.28)} & 
57.90 \scalebox{.7}[1]{(55.72--60.05)} & 
81.55 \scalebox{.7}[1]{(79.79--83.19)} & 
94.75 \scalebox{.7}[1]{(93.68--95.64)} \\
\multirow[l]{6}{=}{\parbox{35mm}{\raggedright Dense\hspace{2.5mm}$\begin{cases}{}\\[40pt]{}\end{cases}$}} &
$k=1$ & 
83.05 \scalebox{.7}[1]{(81.34--84.63)} & 
37.05 \scalebox{.7}[1]{(34.96--39.19)} & 
95.80 \scalebox{.7}[1]{(94.83--96.59)} & 
70.70 \scalebox{.7}[1]{(68.67--72.65)} \\
{} & 
$k=2$ & 
42.30 \scalebox{.7}[1]{(40.15--44.48)} & 
4.75 \scalebox{.7}[1]{(3.90--5.77)} & 
72.05 \scalebox{.7}[1]{(70.04--73.97)} & 
24.75 \scalebox{.7}[1]{(22.91--26.69)} \\
{} & 
$k=3$ & 
12.75 \scalebox{.7}[1]{(11.36--14.28)} & 
0.25 \scalebox{.7}[1]{(0.11--0.58)} & 
31.95 \scalebox{.7}[1]{(29.94--34.03)} & 
4.30 \scalebox{.7}[1]{(3.50--5.28)} \\
{} & 
$k=4$ & 
1.95 \scalebox{.7}[1]{(1.43--2.65)} & 
0.00 \scalebox{.7}[1]{(0.00--0.19)} & 
6.55 \scalebox{.7}[1]{(5.55--7.72)} & 
0.50 \scalebox{.7}[1]{(0.27--0.92)} \\
{} & 
$k=5$ & 
0.10 \scalebox{.7}[1]{(0.03--0.36)} & 
0.00 \scalebox{.7}[1]{(0.00--0.19)} & 
1.00 \scalebox{.7}[1]{(0.65--1.54)} & 
0.00 \scalebox{.7}[1]{(0.00--0.19)} \\
{} & 
$k=6$ & 
0.05 \scalebox{.7}[1]{(0.01--0.28)} & 
0.00 \scalebox{.7}[1]{(0.00--0.19)} & 
0.05 \scalebox{.7}[1]{(0.01--0.28)} & 
0.00 \scalebox{.7}[1]{(0.00--0.19)} \\
\bottomrule
\end{tabular}
\end{center}
{\footnotesize
Estimation used the null space of the canonical coefficients and a stepwise procedure, assessed using the Wilks' statistic, and corrected using a closed testing procedure (\textsc{ctp}). The \textsc{ctp} guarantees that the familywise error rate (\textsc{fwer}) matches the \textsc{pcer} for the first canonical correlation (i.e., for $k=1$).
\par}
\label{tab:results:power+pca}
\end{sidewaystable}

\section{Discussion}

\subsection{Permutation tests}

Compared to univariate, multivariate tests pose the problem of establishing the distributional form for more complicated test statistics; in the parametric case, inference is marred by a set of difficulties: the assumption that all observations are independent and identically distributed following normal theory, the extremely complicated formulas for the density of the canonical correlations, which further depend on the (unknown) population canonical correlations, the sensitivity of asymptotic approximations to departures from assumptions, bias in estimations of parameters, and the validity of these approximations only for particular cases.

Permutation tests address these difficulties in different ways, and their advantages are well known \citep{Ludbrook1998, Nichols2002, Good2005, Pesarin2012}: no underlying distributions need be assumed, non-independence and even heteroscedastic variances can be accommodated, non-random samples can be used, and a wide variety of test statistics are allowed. Moreover, all information needed to build the null distribution lie within the data, as opposed to in some idealised population.

These many benefits extend to inference for multivariate methods. In the case of \textsc{cca}, one benefit is immediately obvious: the complicated formulas and charts for the distribution of the canonical correlations can be bypassed completely, thus with no need to appeal to distributional assumptions. In effect, as shown in Section \ref{sec:results:normality}, even with all variables not following a normal distribution, error rates were still controlled at the nominal level. It should be noted, however, that extremely kurtotic data, such as that generated with a Student's $t$ distribution with extremely low degrees of freedom, caused results to be invalid in the presence of nuisance variables, even with the Huh--Jhun or Theil methods. Such data, however, are rare (recall that with 2 degrees of freedom, the Student's $t$ distribution has infinite variance); most applications of \textsc{cca} investigate datasets that have variables with data that have diverse distributional properties.

Yet, although in the univariate case, algorithms for permutation inference tend to be relatively straightforward to implement and do lead to valid results, for \textsc{cca}, the theory presented in the previous sections and the results with synthetic data show that a simple permutation algorithm that does not consider aspects such as a stepwise estimation of the canonical correlations, nor the inclusion of the null space of canonical coefficients when the two sets of variables do not have the same size, or that does not accommodate specific treatment for nuisance variables, or addresses multiplicity respecting the ordering of the canonical correlations, lead to invalid results.

\subsection{Estimation and multiple testing correction}

Results from Sections \ref{sec:results:estimation} and \ref{sec:results:multiplicity} show that the estimation method that leads to exact, valid results (neither conservative or invalid) is the one that estimates one canonical correlation one at a time, in a stepwise, iterative manner, that includes the null space of the canonical coefficients when the sets of variables have different sizes (i.e., when $P \neq Q$), and that computes adjusted p-values using a closed testing procedure. All alternative approaches led to either invalid or conservative results when considering \textsc{pcer} or the \textsc{fwer}.

It should be emphasised, however, that there are cases in which the na\"{i}ve permutation method, described at the beginning of Section \ref{sec:theory:permutation} remains valid. The method is valid whenever only the first ($k=1$) canonical component is of interest, and there are no nuisance variables or, if there are nuisance variables, those in the left and right side ($\mathbf{Z}$ and $\mathbf{W}$) are completely orthogonal (thus, excluding partial \textsc{cca}). Even though the na\"{i}ve method was not explicitly tested, it is equivalent to the single step method with the null space included, which in the simulations led to an error rate of 4.70\% at test level 0.05 (Table \ref{tab:results:estimation+mtp}). The reason why it remains valid is that, if interest is only in the first canonical component, there is no need to perform an initial \textsc{cca} to allow stepwise removal of previous (before the current $k$ components). Moreover, there is no multiple testing to be considered.

The last column of Table \ref{tab:results:estimation+mtp} may suggest that uncorrected p-values (upper panel) and a \textsc{ctp} (middle) are equivalent for stepwise estimation. They are not, and their differences are manifest in two ways, both previously discussed: first, uncorrected p-values are not monotonically related to the canonical correlations (Figure \ref{fig:pvals}), and second, \textsc{fwer} has potential to be higher than the \textsc{pcer} for $k=1$, even though that did not happen in the simulations.

\subsection{Inference in the presence of nuisance variables}

It is sometimes the case that known, spurious variability needs to be taken into account. For example, variables such as age and sex are often considered confounds. Merely regressing out such nuisance variables from all other variables that are subjected to \textsc{cca}, then proceeding to a simple permutation test, leads to inflated error rates and an invalid test, as expected from Section \ref{sec:theory:nuisance}, and evidenced by the results in Section \ref{sec:results:nuisance}. The dependencies among observations introduced through the residualisation renders the data no longer exchangeable.

This inflated error rate, even after multiple testing correction, is the probably the most striking finding of the current study, as the results can be dramatically affected, particularly if the number of nuisance variables is relatively large compared to the sample size, as shown in Section \ref{sec:results:large}. Transformations that make residuals exchangeable again, through the use of a lower dimensional basis where exchangeability holds, namely, the Huh--Jhun and Theil methods, mitigate the problem, as evidenced by the theory and through the simulations.

Even though both methods led to similarly controlled error rates, they are not equivalent: Huh--Jhun always leads to same canonical components as they would have been obtained from the residualised data, whereas the Theil method can allow for multiple, different solutions depending on the choice of the selection matrix $\mathbf{S}$. \citet{Theil1965} suggested that the choice of the observations to be dropped should consider power; here we suggest that the choice of $\mathbf{S}$ can be based on restrictions on exchangeability: if all original data are freely exchangeable, the Huh--Jhun method is a preferable choice in that it does not require an additional arguments that affect the results; however, it does require a Schur or singular value decomposition of the residual-forming matrix, which is a rank-deficient matrix, such that numerical stability should also be a factor for consideration.

For bipartial \textsc{cca}, while error rates were controlled even in the simple residualisation case, it should be noted that $\mathbf{Z}$ and $\mathbf{W}$ were generated independently in the simulations, such that they were expected to be orthogonal. With real data, possible overlap among columns or linear combinations of columns between $\mathbf{Z}$ and $\mathbf{W}$ create a case that would lie between the two extremes of partial and bipartial \textsc{cca}. In such case, and given the results for partial \textsc{cca}, error rates are not expected to be controlled with simple residualisation. Huh--Jhun and Theil, being able to deal with the most extreme case of dependencies between  $\mathbf{Z}$ and $\mathbf{W}$ (that is, when the two are the same, which define partial \textsc{cca}), constitute a general solution to all cases.

\subsection{Relationship with the \textsc{glm}}

The dangers of residualising both dependent and independent variables in the general linear model (\textsc{glm}) with respect to nuisance variables, then proceeding to a permutation test, as proposed originally by \citet{Kennedy1995} are well known \citep{Anderson2001}. It is not a complete surprise, therefore, that permutation inference for \textsc{cca} would lead to invalid results in similar settings. The original \citet{Huh2001} method \citep[see also][]{Kherad-Pajouh2010} was proposed for the \textsc{glm} as a way to address shortcomings of the Kennedy method in accommodating nuisance variables. Both Kennedy and Huh--Jhun were evaluated by \citet{Winkler2014}: among the methods that can be considered for permutation inference in the \textsc{glm}, Huh--Jhun is the only that cannot be used directly with exchangeability blocks, as the reduction to a lower dimensional space does not respect the block structure. The solution proposed here for permutation inference for \textsc{cca} in the presence of exchangeability blocks, that uses the Theil method, is expected to solve the same problem also for the \textsc{glm}, i.e., as a replacement for Huh--Jhun in cases where the data have a block dependence structure, as it does for freely exchangeable data \citep{Ridgway2009}.

As in the univariate case, permutation tests in the presence of nuisance variables are approximate. Their exactness is in the sense that, under the null hypothesis, the probability of finding a p-value smaller or equal to the test level is the test level itself. Such tests are not perfectly exact as the true relationship between the nuisance variables and the variables of interest are not known and needs be estimated. Even in the absence of nuisance variables, however, permutation tests that use only a fraction of the total number of possible permutations are also approximate, for not covering the whole permutation space (the number of potential permutations tends to be very large, and grows very rapidly with increases in sample size). The same holds for other resampling methods that do not use all possible rearrangements of the data. Regardless of the reason why the tests are approximate, results are known to converge asymptotically to the true p-values.

\subsection{Choice of the statistic}

Among the two test statistics considered, Wilks' ($\lambda_k$) tends to be more powerful than Roy's ($\theta_k$) for effects that span multiple canonical components; the converse holds for signals concentrated in only a few of the canonical components, i.e., when many of the canonical variables are zero; in these cases, Roy's tend to be more powerful than Wilks', as shown in Section \ref{sec:results:statistic}. The respective formulas (Equations \ref{eqn:Wilks} and \ref{eqn:Roy}) give insight on why that is the case: Roy's statistic is invariant to canonical correlations other than the first (largest), whereas Wilks' pool information across all correlations; past simulations, reviewed by \citet{Johnstone2017}, corroborate to the finding.

The use of these two statistics for any canonical correlation other than the first (i.e., for $k > 1$) is possible in the proposed iterative procedure because, for the current position $k$ being tested, all the variance associated with the previous canonical components at positions $\{1, \ldots, k-1\}$ will have already been removed from the model (Sections \ref{sec:theory:permutation} and \ref{sec:algorithm}), such that the largest canonical correlation (Roy's statistic) is the current one being tested; for Wilks', the procedure holds because these earlier canonical correlations are not marked as zero; instead, they are ignored altogether when the statistic is computed, as if the previous canonical components have never existed.

Wilks' lambda and Roy's largest root are not the only possible statistics that can be considered for \textsc{cca}, and permutation tests allow the use of yet others. Some, such as Hotelling--Lawley and Pillai--Bartlett, were considered by \citet{Friederichs2003}. Using simulations and Monte Carlo results, the authors found that parametric distributions of these classical multivariate statistics were accurate, and could be obtained quickly at low computational cost; it should be noted, however, that the study used normally distributed simulated data, in which case parametric assumptions are known to hold.

\subsection{Relationship with previous studies}

While a number of studies have used permutation tests with \textsc{cca}, not many investigated the performance of these tests. \citet{Nandy2003} proposed a non-parametric strategy for inference with \textsc{cca} for the investigation of task-based f\textsc{mri} time series: the method uses a resting-state (no task-related activity) dataset to build the null distribution; as resampling time series can be challenging due to temporal autocorrelation, the null distribution uses multiple voxels selected far apart from each other so as to also avoid issues with spatial correlation. The approach differs from the one presented here in that it uses subject-level time series (as opposed to between-subject analyses), is specific to brain imaging (the proposed method is general) and does a resampling method that shares similarities with, yet is not the same as permutation. \citet{Eklund2011} specifically used permutation tests for \textsc{cca} with f\textsc{mri} time series whitened with a combination of methods to allow permutation; the authors demonstrated that permutation tests for both the \textsc{glm} and \textsc{cca} could be greatly accelerated through the use of graphics processing units (\textsc{gpu}s).

\citet{Kazi-Aoual1995} proposed analytical formulas for the first three moments of the permutation distribution of Pillai's trace for \textsc{cca}; these moments can be used to fit a \citet{Pearson1895} type \textsc{iii} distribution, from which p-values can be obtained. \citet{Legendre2011} studied parametric and permutation tests for redundancy analysis \cite[\textsc{rda};][]{Rao1964} and for canonical correspondence analysis \cite[referred to also by the acronym ``\textsc{cca}'';][]{terBraak1986}; the authors found that a simultaneous test of all canonical eigenvalues for the respective axes (eigenvectors of predicted response variables in a linear model) in \textsc{rda}, despite simple, is not valid, whereas a marginal test on each eigenvalue, as well as a ``forward'' test in which previously tested canonical axes are added to a matrix of nuisance variables, performs well, even if conservatively for axes other than the first. \citet{Yoo2012} investigated the relationship between \textsc{cca} and regression, proposing the use of permutations and studying cases without nuisance variables; in the method, for the $k$-th canonical correlation, variance not already explained by canonical variables $\{1,\ldots,k\}$ in one of the sides is permuted, whereas the other variables remain fixed. \citet{Turgeon2018} considered using a small number of permutations for \textsc{cca}, recording of the empirical distribution function, then using it to estimate the parameters of a Tracy--Widom distribution \citep{Tracy1996, Johnstone2008} for cases in which the number of observations is smaller than the number of variables in either $\mathbf{Y}$ or $\mathbf{X}$; the distribution is then used to obtain p-values; data are assumed to follow a normal distribution, and inference is for the largest canonical correlation.

Permutation tests for the method of partial least squares \citep[\textsc{pls};][]{Tucker1958, Wold1983, McIntosh1996, McIntosh2004} have been considered. For example, \citet{Chin2010} and \citet{Sarstedt2011} used a permutation test to investigate how differences in the strength of association between variables (magnitude of estimates) further differed between two or more groups. These would be equivalent to, in the context of \textsc{cca}, testing whether canonical correlations obtained across different groups would differ. \citet{LeFloch2012} investigated strategies for dimensionality reduction and regularisation for imaging and genetic data, whereas \citet{Grellmann2015} compared the performance of variants of \textsc{cca} and \textsc{pls} for similar problems. Both studies used direct permutation of the data, and were mostly focused on the relative performance of the different methods, offering no specific treatment of nuisance variables or the other aspects considered here, and which concern validity. \citet{Monteiro2016} investigated a strategy for sparse \textsc{pls} and sparse \textsc{cca} in which data are split into training and hold out, and inference uses permutation of the training data, with coefficients applied to test data, in which measurements of association are computed.

The current paper therefore fills a substantial knowledge gap, whereby not many studies considered at all the validity of permutation inference with \textsc{cca}, but those that did approach the topic were not sufficiently general; none covered the topics discussed here. Moreover, in principle, the method as proposed can be used with subject-level f\textsc{mri} or other timeseries data provided that whitening has been successful in removing temporal dependencies. Additionally, given the conceptual similarity between \textsc{pls} and \textsc{cca}, it is possible that permutation inference for \textsc{pls} would require similar strategies as described in Section \ref{sec:theory}, particularly in the presence of nuisance variables and re-use of variance already explained. Whether that is the case, it is a question that remains open for future investigation.

\subsection{Recommendations}

Given the above results, the main recommendations for permutation inference for \textsc{cca} can be summarised as follows:

\begin{itemize}
\item When studying a given $k$-th canonical variable or canonical correlation, $1 < k \leqslant K$, remove the effects of the previous ones, i.e., the variance from one set that has already been explained by the other, as represented by the earlier canonical variables. These effects are surely significant (regardless of the test level), otherwise the current canonical variable or correlation would not be under consideration. Ignoring the earlier ones cause the error rates be inflated (empirical evidence provided in Section \ref{sec:results:estimation}).
\item For sets of variables with different sizes (i.e., $P \leqslant Q$), ensure that the variability not represented by the canonical variables produced at the first permutation is considered in all and every permutation. That is, include the null space of the canonical coefficients when computing the variables subjected to permutation. Not including the null space leads to excess false positives (empirical evidence provided in Section \ref{sec:results:estimation}).
\item Do not use simple p-values for inference, and make sure that a closed testing procedure is used. Using simple, uncorrected p-values has two negative consequences: (i) both the \textsc{pcer} and \textsc{fwer} are inflated, and (ii) since simple p-values are not guaranteed to be monotonically related to the canonical correlations, the resulting test is inadmissible (empirical evidence provided in Section \ref{sec:results:multiplicity}).
\item For the same reason, do not use \textsc{fdr} to correct for multiple testing after using simple p-values: while the p-values themselves satisfy the requirements of \textsc{fdr}, they lead to an inadmissible test even after correction, leading to non-sensical results whereby a stronger canonical correlation may be less significant than a weaker one (empirical evidence provided in Section \ref{sec:results:multiplicity}).
\item While valid, inference using the distribution of the maximum statistic across canonical correlations leads do conservative results, except for the first canonical correlation (empirical evidence provided in Section \ref{sec:results:multiplicity}).
\item If regressing out nuisance variables from both sets of variables subjected to \textsc{cca}, make sure that the residuals are transformed to be exchangeable, e.g., with the Huh--Jhun or Theil methods, then permuted accordingly. Failure to observe this recommendation leads to excess false positives, particularly when the number of nuisance variables is a large fraction of the sample size (empirical evidence provided in Sections \ref{sec:results:nuisance} and \ref{sec:results:large}).
\end{itemize}

All these recommendations are integrated into Algorithm 1.


\section{Conclusion}

As evidenced by the theory and simulations in the previous sections, a simple permutation procedure leads to invalid results: (i) simple p-values are not admissible for inference in \textsc{cca}, lead to excess \textsc{pcer} and \textsc{fwer}, and cannot be corrected using generic methods based on p-values such as \textsc{fdr}; (ii) ignoring the variability already explained by previous canonical variables leads to inflated error rates for all canonical correlations except for the first; (iii) regression of the same set of nuisance variables from both sides of \textsc{cca} without further consideration leads to inflated error rates; and (iv) the classical method for multiple testing correction, that uses the distribution of the maximum statistic leads to conservative results. The use of a stepwise estimation procedure, transformation of the residuals to a lower dimensional basis where exchangeability holds, and correction for multiple testing via closure, ensures the validity of permutation inference for \textsc{cca}.


\appendix

\section{Ancillary functions}

Algorithm 1 requires two relevant ancillary functions: one to compute the semi-orthogonal matrix $\mathbf{Q}$, and another to conduct the \textsc{cca} proper and obtain the canonical coefficients $\mathbf{A}$ and $\mathbf{B}$; these two functions are described in pseudo-code in Algorithms 2 and 3. The ``semiortho'' function takes as input a residual-forming matrix $\mathbf{R}$ and, optionally, a selection matrix $\mathbf{S}$. If $\mathbf{S}$ is supplied, it computes $\mathbf{Q}$ using the Theil method; otherwise, it uses the Huh--Jhun method (Table \ref{tab:semiortho}).

As shown, ``semiortho'' uses Schur decomposition for Huh--Jhun, but that decomposition can be replaced by singular value decomposition (\textsc{svd}) or \textsc{qr} decomposition. Another possibility consists of never using $\mathbf{R}$ directly, computing instead an orthogonal basis for the null space of $\mathbf{Z}$ (not shown; it would require taking $\mathbf{Z}$ as an input argument). All these are expected to produce the same results. However, as the residual-forming matrix is rank deficient and idempotent, all its eigenvalues are identical to 0 or 1. Thus, considerations about numerical stability and float point arithmetic \citep{Moler2004}, as well as speed, should determine the best choice for a particular programming language or computing architecture.

\begin{myalgorithm}
\singlespacing
\noindent
\caption{The ``\textbf{semiortho}'' function, used in Algorithm 1.}
\HRule
\vspace{3mm}
{\small
\begin{algorithmic}[1]
\Require $\mathbf{R}_{N \times N}$. \textbf{Optional:} $\mathbf{S}_{N \times N'}$.
\Comment{Inputs.}
\If{$\text{exist}(\mathbf{S})$}
\Comment{If a selection matrix was supplied.}
\State $\mathbf{Q}_{N \times N'} \leftarrow \mathbf{R}\mathbf{S'}(\mathbf{S}\mathbf{R}\mathbf{S}')^{-1/2}$
\Comment{Use Theil.}
\Else 
\Comment{Otherwise, use Huh--Jhun.}
\State $\mathbf{Q}_{N \times N},\mathbf{E} \leftarrow \text{schur}(\mathbf{R})$
\Comment{Schur decomposition.}
\State $\mathbf{Q}_{N \times N'} \leftarrow [\,\mathbf{q}_{n}\,|\,e_n > 0\,]$
\Comment{Keep columns $\mathbf{q}_{n}$ with positive eigenvalues $e_n$.}
\EndIf
\State \Return $\mathbf{Q}$
\Comment{Return the semi-orthogonal matrix.}
\end{algorithmic}}
\noindent
\HRule\\[2mm]
\setstretch{\lspac}
\end{myalgorithm}

\begin{myalgorithm}
\singlespacing
\noindent
\caption{The ``\textbf{cca}'' function, used in Algorithm 1.}
\HRule
\vspace{3mm}
{\small
\begin{algorithmic}[1]
\Require $\mathbf{Y}_{N \times P}, \mathbf{X}_{N \times Q}, R, S$.
\Comment{Inputs.}
\State $K \leftarrow \min (P,Q)$
\Comment{Number of canonical components.}
\State $\mathbf{Q}_{\mathbf{Y}},\mathbf{R}_{\mathbf{Y}},\mathbf{T}_{\mathbf{Y}}' \leftarrow \text{qr}(\mathbf{Y})$;
\Comment{\textsc{qr} decomposition.}
\State $\mathbf{Q}_{\mathbf{X}},\mathbf{R}_{\mathbf{X}},\mathbf{T}_{\mathbf{X}}' \leftarrow \text{qr}(\mathbf{X})$;
\Comment{\textsc{qr} decomposition.}
\State $[\mathbf{L}_{P \times P},\mathbf{D},\mathbf{M}_{Q \times Q}'] \leftarrow \text{svd}(\mathbf{Q}_{\mathbf{Y}}'\mathbf{Q}_{\mathbf{X}})$;
\Comment{Singular value decomposition.}
\State{$[r_1, \ldots, r_K] \leftarrow [d_{11}, \ldots, d_{KK}]$}
\Comment{Canonical correlations (diagonal of $\mathbf{D}$).}
\State $\mathbf{L}_{P \times K} \leftarrow [\,\mathbf{l}_{k}, \,k=\{1,\ldots,K\}\,]$
\Comment{Retain the first $K$ columns of $\mathbf{L}$.}
\State $\mathbf{M}_{Q \times K} \leftarrow [\,\mathbf{m}_{k}, \,k=\{1,\ldots,K\}\,]$
\Comment{Retain the first $K$ columns of $\mathbf{M}$.}
\State $\mathbf{A}_{P \times K} \leftarrow \mathbf{T}_{\mathbf{Y}}\mathbf{R}_{\mathbf{Y}}^{-1}\mathbf{L}\sqrt{N-R}$
\Comment{Canonical coefficients, left side.}
\State $\mathbf{B}_{Q \times K} \leftarrow \mathbf{T}_{\mathbf{X}}\mathbf{R}_{\mathbf{X}}^{-1}\mathbf{M}\sqrt{N-S}$
\Comment{Canonical coefficients, right side.}
\State \Return $\mathbf{A}, \mathbf{B}, [r_1, \ldots, r_K]$
\Comment{Return coefficients and correlations.}
\end{algorithmic}}
\noindent
\HRule\\[2mm]
\setstretch{\lspac}
\end{myalgorithm}

The ``cca'' function takes as main inputs the sets of variables $\mathbf{Y}$ and $\mathbf{X}$. These will have been mean-centered and possibly residualised outside the function, such that no further mean-centering or residualisation is performed; if mean-centering was performed, then, at a minimum, the other two arguments are $R=S=1$; if other variables were regressed out, as in part, partial, or bipartial \textsc{cca}, then $R$ and $S$ are supplied with their corresponding values. The algorithm uses the method described by \citet{Bjorck1973}, and is based on results of \citet{Olkin1951} and \citet{Golub1969}; additional details can be found in \citet{Seber1984}. Inside this function, variables $\mathbf{Q}$ and $\mathbf{R}$ (subscripts omitted) refer to the factors of a \textsc{qr} factorization, hence with a different meaning than the similarly named matrices used elsewhere this paper. In the algorithm, $\mathbf{Y}$ and $\mathbf{X}$ are subjected to \textsc{qr} decomposition with pivoting (hence the matrices $\mathbf{T}$, subscripts omitted), using a numerically stable Householder transformation \citep{Golub2013}. The inner product $\mathbf{Q}_{\mathbf{Y}}'\mathbf{Q}_{\mathbf{X}}$ of the orthogonal matrices from \textsc{qr} is subjected to singular value decomposition; the diagonal elements of $\mathbf{D}$ are the canonical correlations (line 5). The remaining computations are for the canonical coefficients: these are obtained via back substitution by solving the triangular sets of equations $\mathbf{L}=\mathbf{R}_{\mathbf{Y}}^{-1}\mathbf{A}$ and $\mathbf{M}=\mathbf{R}_{\mathbf{X}}^{-1}\mathbf{B}$. The permutation matrices $\mathbf{T}_{\mathbf{Y}}$ and $\mathbf{T}_{\mathbf{X}}$ are used for reordering. The constant factors in the square roots are normalising scalars to ensure unit variance for the canonical variables $\mathbf{U}$ and $\mathbf{V}$ (not returned by the algorithm, but computable as $\mathbf{U}=\mathbf{Y}\mathbf{A}$ and $\mathbf{V}=\mathbf{X}\mathbf{B}$, Equation \ref{eqn:UV}); if an intercept was explicitly included in $\mathbf{Z}$ and $\mathbf{W}$ (for bipartial), these constant factors are as shown; if instead the data were mean-centered, further subtract 1 before taking the square root. Regardless, omission of these constant terms do not affect the canonical correlations.

{\footnotesize
\section*{Source code}
\noindent
Code related to this paper is available at \href{https://github.com/andersonwinkler/PermCCA}{https://github.com/andersonwinkler/PermCCA}.
\par}

{\footnotesize
\section*{Acknowledgements}
\noindent
The authors thank Drs.\ Julia O.\ Linke and Daniel S.\ Pine (National Institutes of Health, \textsc{nih}) for the invaluable discussions. A.M.W. receives support through the \textsc{nih} Intramural Research Program (ZIA-MH002781 and ZIA-MH002782). T.E.N. received support from the Wellcome Trust, 100309/Z/12/Z. This work utilized computational resources of the \textsc{nih} \textsc{hpc} Biowulf cluster (\href{https://hpc.nih.gov}{https://hpc.nih.gov}).
\par}

{\footnotesize
\setlength{\bibsep}{0.0pt}
\begin{spacing}{0.9}
\bibliographystyle{elsarticle-harv}
\bibliography{references.bib}
\end{spacing}}

\end{document}